\documentclass[referee]{aastex63}

\RequirePackage{xspace}
\RequirePackage{amsmath}
\RequirePackage{amsfonts}
\RequirePackage{amssymb}

\def\ifundefined#1{\expandafter\ifx\csname#1\endcsname\relax}

\makeatletter
\newcommand*{\rom}[1]{\expandafter\@slowromancap\romannumeral #1@}
\makeatother

\def\la{\mathrel{\hbox{\rlap{\hbox{\lower4pt\hbox{$\sim$}}}\hbox{$<$}}}}
\def\ga{\mathrel{\hbox{\rlap{\hbox{\lower4pt\hbox{$\sim$}}}\hbox{$>$}}}}

\newcommand{\be}{\begin{equation}}
\newcommand{\ee}{\end{equation}}

\newcommand{\bea}{\begin{eqnarray}}
\newcommand{\eea}{\end{eqnarray}}

\ifundefined{ensuremath}\def\ensuremath#1{\relax\ifmmode{#1}}
\else${#1}$\fi\else\relax\fi
\ifundefined{nuc}\def\nuc#1#2{\relax\ifmmode{}^{#1}{\protect\text{#2}}
\else${}^{#1}$#2\fi}\else\relax\fi

\ifundefined{ion}
\newcommand\ion[2]{#1$\;${\ifx\@currsize\normalsize\small \else
\ifx\@currsize\small\footnotesize \else
\ifx\@currsize\footnotesize\scriptsize \else
\ifx\@currsize\scriptsize\tiny \else
\ifx\@currsize\large\normalsize \else
\ifx\@currsize\Large\large
\fi\fi\fi\fi\fi\fi
\rmfamily\rom{#2}}\relax}\else\relax\fi

\newcommand{\etal}{et al.\xspace}

\newcommand{\kmps}{\ensuremath{\text{km}~\text{s}^{-1}}\xspace}

\makeatletter

\newcommand\Autoref[1]{\@first@ref#1,@}
\def\@throw@dot#1.#2@{#1}\def\@set@refname#1{    \edef\@tmp{\getrefbykeydefault{#1}{anchor}{}}    \def\@refname{\@nameuse{\expandafter\@throw@dot\@tmp.@autorefname}s}}
\def\@first@ref#1,#2{  \ifx#2@\autoref{#1}\let\@nextref\@gobble  \else    \@set@refname{#1}    \@refname~\ref{#1}    \let\@nextref\@next@ref  \fi  \@nextref#2}
\def\@next@ref#1,#2{   \ifx#2@ and~\ref{#1}\let\@nextref\@gobble   \else, \ref{#1}   \fi   \@nextref#2}

\makeatother
\xspaceaddexceptions{]}

\newcommand{\vsi}{\ensuremath{v_\text{Si II}}\xspace}
\newcommand{\Bline}{\ion{Si}{2}~$\lambda5972$\xspace}
\newcommand{\Rline}{\ion{Si}{2}~$\lambda6355$\xspace}
\newcommand{\polinplot}{$M_B$-vs-$v_\text{Si II}$\xspace}
\newcommand{\BV}{$(B_\text{max}-V_\text{max})_0$\xspace}

\graphicspath{{./figs/}{./figs/GMM_branch/}{./figs/GMM_v_p5_p6/}{./figs/GMM_m_p5_p6/}{./figs/spextractor_plots/}{./figs/GMM_m_v_p5_p6/}{./figs/gmm_polin/}}

\begin{document}

\title{\textit{Carnegie Supernova Project:} Classification of Type Ia
               Supernovae}

\author[0000-0002-5380-0816]{Anthony Burrow}
\affiliation{Homer L.~Dodge Department of Physics and Astronomy, University of
             Oklahoma, Rm 100 440 W. Brooks, Norman, OK 73019-2061}

\author[0000-0001-5393-1608]{E.~Baron}
\affiliation{Homer L.~Dodge Department of Physics and Astronomy, University of
             Oklahoma, Rm 100 440 W. Brooks, Norman, OK 73019-2061}
\affiliation{Hamburger Sternwarte, Gojenbergsweg 112, 21029 Hamburg, Germany}

\author[0000-0002-5221-7557]{Chris Ashall}
\affiliation{Department of Physics, Florida State University, Tallahassee, FL
             32306, USA}

\author[0000-0003-4625-6629]{Christopher~R.~Burns}
\affiliation{Observatories of the Carnegie Institution for Science, 813 Santa
             Barbara St., Pasadena, CA 91101, USA}

\author[0000-0002-4338-6586]{N.~Morrell}
\affiliation{Las Campanas Observatory, Carnegie Observatories, Casilla 601, La
             Serena, Chile}

\author[0000-0002-5571-1833]{Maximilian~D.~Stritzinger}
\affiliation{Department of Physics and Astronomy, Aarhus University, Ny
             Munkegade 120, DK-8000 Aarhus C, Denmark.}

\author[0000-0001-6272-5507]{Peter J.~Brown}
\affiliation{George P. and Cynthia Woods Mitchell Institute for Fundamental
             Physics and Astronomy, Department of Physics and Astronomy, Texas
             A\&M University, College Station, TX 77843, USA}

\author[0000-0001-5247-1486]{G.~Folatelli}
  \affiliation{Instituto de Astrof\'isica de La Plata (IALP), CONICET,
    Paseo del Bosque S/N, B1900FWA La Plata, Argentina}
\affiliation{Facultad de Ciencias Astron\'omicas y Geof\'isicas
    Universidad Nacional de La Plata, Paseo del Bosque, B1900FWA, La
    Plata, Argentina}

\author[0000-0003-3431-9135]{Wendy L.~Freedman}
\affiliation{Department of Astronomy \& Astrophysics, University of Chicago, 5640 South Ellis Avenue, Chicago, IL 60637, USA}

\author[0000-0002-1296-6887]{Llu{\'i}s Galbany}
\affiliation{Departamento de F\'isica Te\'orica y del Cosmos,
  Universidad de Granada, E-18071 Granada, Spain}

\author[0000-0002-4338-6586]{P.~Hoeflich}
\affiliation{Department of Physics, Florida State University, Tallahassee, FL
             32306, USA}

\author[0000-0003-1039-2928]{Eric~Y.~Hsiao}
\affiliation{Department of Physics, Florida State University, Tallahassee, FL
             32306, USA}

\author[0000-0002-6650-694X]{Kevin Krisciunas}
\affiliation{George P. and Cynthia Woods Mitchell Institute for Fundamental
             Physics and Astronomy, Department of Physics and Astronomy, Texas
             A\&M University, College Station, TX 77843, USA}

\author[0000-0003-2734-0796]{M.~M.~Phillips}
\affiliation{Las Campanas Observatory, Carnegie Observatories, Casilla 601, La
             Serena, Chile}

\author[0000-0001-6806-0673]{Anthony~L.~Piro}
\affiliation{Observatories of the Carnegie Institution for Science, 813 Santa
             Barbara St., Pasadena, CA 91101, USA}

\author[0000-0002-8102-181X]{Nicholas~B.~Suntzeff}
\affiliation{George P. and Cynthia Woods Mitchell Institute for Fundamental
             Physics and Astronomy, Department of Physics and Astronomy, Texas
             A\&M University, College Station, TX 77843, USA}

\author[0000-0002-0786-7307]{Syed Uddin}
\affiliation{Observatories of the Carnegie Institution for Science, 813 Santa
             Barbara St., Pasadena, CA 91101, USA}

\submitjournal{ApJ}

\received{\today}
\revised{\today}
\accepted{\today}

\correspondingauthor{Anthony Burrow}
\email{anthony.r.burrow-1@ou.edu}

\begin{abstract}

We use the spectroscopy and homogeneous photometry of 97 Type Ia supernovae
obtained by the \emph{Carnegie Supernova Project} as well as a subset of 36
Type Ia supernovae presented by \citet{Zheng_IaLCIII18} to examine
maximum-light correlations in a four-dimensional (4-D) parameter space:
$B$-band absolute magnitude, $M_B$, \Rline velocity, \vsi, and \ion{Si}{2}
pseudo-equivalent widths pEW(\Rline) and pEW(\Bline). It is shown using
Gaussian mixture models (GMMs) that the original four groups in the Branch
diagram are well-defined and robust in this parameterization. We find three continuous
groups that describe the behavior of our sample in [$M_B$, \vsi]
space.
Extending the GMM into the full 4-D
space yields a grouping system that only slightly alters group definitions in
the [$M_B$, \vsi] projection, showing that most of the clustering information
in [$M_B$, \vsi] is already contained in the 2-D GMM groupings.
However, the full 4-D space does divide group membership for faster
objects between core-normal and broad-line objects in the Branch
diagram. A significant correlation between
$M_B$ and pEW(\Bline) is found, which implies that Branch group membership can
be well-constrained by spectroscopic quantities alone. In general, we find that
higher-dimensional GMMs reduce the uncertainty of group
membership for objects between the originally defined Branch groups. We also
find that the broad-line Branch group becomes nearly distinct with the
inclusion of \vsi, indicating that this subclass of SNe Ia may be somehow
different from the other groups.

\end{abstract}

\section{Introduction}

Type Ia supernovae (SNe~Ia) are  extragalactic distance indicators used to
measure the expansion history of the universe. However, it is not clear what
their progenitor systems and explosion mechanisms are, and it may be that
SNe~Ia arise from multiple progenitor and explosion channels.
Time-domain astronomy has reached an era of larger data and
advanced statistical analysis.  With more information, new empirical groups and
correlations can be identified to try to understand the nature of time-variable
objects such as SNe~Ia. We apply modern data analysis methods to the
homogeneous photometric and spectroscopic sample from the
\textit{Carnegie Supernova Project} (CSP) I+II
\citep[][N.~Morrell et al, in prep.; N.~Suntzeff et al, in prep; S.~Uddin, in prep]{folatelli13,kkcsp1,HsiaoCSPII19,PhilCSPII19}
in order to study established correlations and attempt to quantify their
identities.

SNe~Ia have been made correctable candles based on their light curve shape
using a number of empirical parameters, such as $\Delta m_{15}(B)$ \citep{philm15,rpk96,hametal96a,philetal99},
stretch $s$ \citep{goldhetal01}, and the color stretch parameter $s_{BV}$
\citep{Burns_CSP14}. The rate of decline of the light curve postmaximum was
first associated with peak luminosity through the Phillips relation,
which asserts that fast decliners are dimmer than slower decliners
\citep{philm15}. However, it was quickly realized that intrinsically red
supernovae are also intrinsically dim \citep{tripp98}.
It has long been understood that the light curve shape relation does
not capture all of the diversity in the SNe~Ia sample. Several attempts have
been made to use photometric and spectroscopic indicators to sort supernovae in
a different dimension \citep[e.g.][]{nugseq95,bongard06a,bailey09,Ashall20}.
\citet{benetti05} used the velocity gradient of the \Rline line to divide
SNe~Ia into three classes, and \citet{xwang09} used the velocity of the same
line at maximum to divide them into two classes. \citet{branchcomp206} compared
the pseudo-equivalent width of the \Bline line with that of the \Rline line to
define four subclasses of SNe Ia: core-normal, shallow-silicon, broad-line, and
cool. Henceforth, we will refer to these subclasses as the Branch groups.

Here, we use the optical spectral data from the CSP I+II to study spectroscopic
classification of SNe~Ia. Recently, \citet{Zheng_IaLCIII18} suggested an
empirical fitting method for SN~Ia light curves using the risetime and the
\Rline velocity at maximum light as a method for calibrating the $B$-band peak
luminosity. Using the results of \citet{Zheng_IaLCIII18}, \citet{polin19hedet}
plotted peak absolute magnitude in the $B$-band versus \Rline velocity at maximum light.
This plot, which we henceforth refer to as the \polinplot diagram, seemed to
indicate a dichotomy in supernova explosion mechanisms which, when
displayed so as to indicate $B_\text{max}-V_\text{max}$
color that has been corrected for Milky Way (MW) Galaxy extinction, led to
an interpretation that the red supernovae could be well fit by explosion models
of helium detonations on sub-Chandrasekhar mass progenitors
\citep{polin19hedet}. Our results show that when intrinsic \BV color
is used, we do not find a break between supernovae into a primary group
consisting mostly of bluer members and a secondary group with mostly redder
members. Thus, the \polinplot diagram does not appear to cleanly delineate
between Chandrasekhar mass and sub-Chandrasekhar mass explosions.

As a guide to the reader, we summarize the definition of the seven
  different groups we identify in this work. The Branch diagram
  \citep{branchcomp206} is a plot of the pseudo-equivalent width (pEW) of
  \Bline versus the pEW of \Rline, both at maximum light in the
  $B-$band. This diagram
  was found by \citet{branchcomp206} to cluster into
  four groups:
  shallow-silicons (with low values of pEW for both lines, so in the
  lower left corner of the diagram); core-normals, roughly in the
  central part of the diagram; cools, with moderate values of
  pEW(\Rline), but large values of pEW(\Bline), so occupying the upper
  middle portion of the diagram; and broad-lines with high values of
  pEW(\Rline) and spanning the upper range of pEW(\Bline), so
  occupying the right-hand portion of the diagram. The \polinplot diagram
  \citep{polin19hedet} is a plot of $M_B$ (determined from a model)
  versus \vsi, the velocity of the \Rline at maximum $B$ light. In this
  diagram we find three groups: Main, near normal brightness with $M_B
  \sim -19.5$ and velocities $\vsi \la 12,000$~\kmps; Dim, with
  $M_B \ga -18.8$ and velocities $\vsi \la 12,000$~\kmps; and Fast, with $M_B
  \sim -19.5$ (having a larger spread than the Main group) and velocities
  $\vsi \ga 12,000$~\kmps.

\section{Data}
\label{sec:data}

We include supernovae whose time of maximum is known, whose $M_B$ (the peak
absolute magnitude in the $B$-band at maximum light) has been determined
\citep[][S.~Uddin, \etal, in prep]{Burns_CSP14,Burns_CSP18}, and whose
spectra fall into the epoch range $| t_{epoch} |< 7$ days past the inferred
time of $B$ maximum. From the initial 364 SNe in the CSP I+II set,
our sample consists of a subset of 97 objects. We include additional
supernovae available using data from \citet{Zheng_IaLCIII18}. In doing so,
there is a deviation from a completely homogeneous sample of data, however we
include this data set not only to increase the sample size, but to compare our
results with those of \citet{polin19hedet} in
\autoref{sec:polin-plot-comparison} and beyond. From the
\citet{Zheng_IaLCIII18} set of 54 SNe, 14 are shared with CSP I,
and CSP I spectra were preferred. We include 36 unique SNe from the
\citet{Zheng_IaLCIII18} sample. In total, we study a sample of 133
objects.

We examine spectroscopic classifications using the CSP I+II samples. The values
of the photometric quantities, for example, $m_{B, \text{max}}$,
$s_{BV}$, host $A_V$, etc. were computed using SNooPy \citep{Burns_CSP14}, which include
K-corrections. $M_B$ is then determined from the value of $m_{B, \text{max}}$
determined by SNooPy, the MW extinction, the SNooPy-inferred host extinction,
and the distance modulus using $H_0 = 72$~\kmps~Mpc$^{-1}$. The redshift distribution of
the CSP I SNe is $0.0037 \le z \le 0.0835$ \citep{kkcsp1}, and that of the
CSP II SNe is $0.03 \le z \le 0.10$ \citep{PhilCSPII19}. $M_B$ and $s_
{BV}$ values for the CSP I+II sample used in this analysis are provided in
\autoref{tab:full_data}.

We use values of $M_B$ that are corrected for MW Galaxy extinction
and host galaxy extinction directly from \citet{Zheng_IaLCIII18} Table 1 with
no remeasurement. These photometric values from \citet{Zheng_IaLCIII18} are
supplemented with spectroscopic information from the
Open Supernova Catalog\footnote{\url{https://sne.space}}
\citep{guillochon_osc17}. The spectral epoch for the
measurement is always chosen to be the epoch that is the closest available
to the time of maximum. The values of $M_B$ determined by \citet{Zheng_IaLCIII18}
were made with $H_0 = 70$~km~s$^{-1}$~Mpc$^{-1}$ and host extinction
was estimated using
MLCSk2 fitting \citep{JRK07} with $R_V$ held fixed at a value of  $1.8$.  Their sample has
a maximum redshift of $z=0.039$, but excluded all SNe with
host $E(B-V) > 0.3$~mag. K-corrections were not included. For
CSP I SNe in the \citet{Zheng_IaLCIII18} sample,
we use the SNooPy determinations. The values used from the unique
\citet{Zheng_IaLCIII18} objects included in this analysis are also provided in
the bottom section of \autoref{tab:full_data}.

\startlongtable
\begin{deluxetable*}{lDccccc}
    \tablecaption{Spectroscopic and photometric information for all 133 SNe
                  from both the CSP I+II samples and the \citet{Zheng_IaLCIII18}
                  sample. Values from the \citet{Zheng_IaLCIII18} sample are
                  given in the bottom sector of this table.
        \label{tab:full_data}}
    \tablewidth{0pt}
    \tablehead{
        \colhead{SN} & \multicolumn2c{Epoch (days} & \colhead{\vsi} & \colhead{pEW(\Bline)}
            & \colhead{pEW(\Rline)} & \colhead{$M_B$} & \colhead{$s_{BV}$} \\
        \colhead{} & \multicolumn2c{past $B_\text{max}$)} & \colhead{(1000 km s$^{-1}$)}
            & \colhead{(\AA)} & \colhead{(\AA)} & \colhead{(mag)} & \colhead{}
    }
    \decimals
    \startdata
        ASASSN-14hr & 5.3 & 13.6 $\pm$ 0.3 & 30.5 $\pm$ 4.2 & 103.1 $\pm$ 5.3 & -19.10 $\pm$ 0.12 & 0.79 $\pm$ 0.05 \\
        ASASSN-14hu & 6.9 & 12.5 $\pm$ 0.2 & 7.4 $\pm$ 1.2 & 86.7 $\pm$ 2.8 & -19.52 $\pm$ 0.12 & 1.08 $\pm$ 0.05 \\
        ASASSN-14kq & -2.3 & 10.5 $\pm$ 0.3 & 10.7 $\pm$ 1.5 & 82.5 $\pm$ 3.5 & -19.44 $\pm$ 0.11 & 1.19 $\pm$ 0.05 \\
        ASASSN-14mf & 5.2 & 9.9 $\pm$ 0.3 & 23.8 $\pm$ 2.0 & 103.8 $\pm$ 3.8 & -19.45 $\pm$ 0.15 & 0.98 $\pm$ 0.05 \\
        ASASSN-14my & 3.6 & 12.0 $\pm$ 0.3 & 26.1 $\pm$ 3.9 & 106.9 $\pm$ 4.3 & -19.47 $\pm$ 0.12 & 0.91 $\pm$ 0.05 \\
        ASASSN-15al & 5.9 & 11.7 $\pm$ 0.1 & 19.7 $\pm$ 4.2 & 81.7 $\pm$ 5.5 & -19.21 $\pm$ 0.16 & 1.10 $\pm$ 0.06 \\
        ASASSN-15ba & 4.0 & 11.4 $\pm$ 0.2 & 16.2 $\pm$ 1.7 & 117.3 $\pm$ 3.3 & -19.31 $\pm$ 0.12 & 0.97 $\pm$ 0.05 \\
        ASASSN-15be & 1.4 & 11.8 $\pm$ 0.4 & 3.8 $\pm$ 1.8 & 94.3 $\pm$ 3.8 & -19.37 $\pm$ 0.12 & 1.20 $\pm$ 0.05 \\
        ASASSN-15bm & 1.2 & 10.6 $\pm$ 0.2 & 9.7 $\pm$ 2.6 & 72.3 $\pm$ 4.9 & -19.57 $\pm$ 0.14 & 1.00 $\pm$ 0.05 \\
        ASASSN-15dd & 2.1 & 10.8 $\pm$ 0.3 & 21.0 $\pm$ 1.6 & 102.1 $\pm$ 2.7 & -19.63 $\pm$ 0.15 & 0.83 $\pm$ 0.05 \\
        ASASSN-15fr & 0.2 & 11.7 $\pm$ 0.2 & 20.7 $\pm$ 3.7 & 107.4 $\pm$ 6.1 & -19.21 $\pm$ 0.09 & 0.88 $\pm$ 0.05 \\
        ASASSN-15ga & 5.2 & 9.4 $\pm$ 0.2 & 52.4 $\pm$ 5.1 & 126.1 $\pm$ 5.0 & -17.95 $\pm$ 0.40 & 0.50 $\pm$ 0.06 \\
        ASASSN-15hf & 1.1 & 11.1 $\pm$ 0.3 & 21.3 $\pm$ 2.9 & 83.6 $\pm$ 2.9 & -19.07 $\pm$ 0.38 & 0.91 $\pm$ 0.05 \\
        CSP14acl & 5.0 & 10.1 $\pm$ 0.3 & 16.2 $\pm$ 3.2 & 116.0 $\pm$ 5.4 & -18.91 $\pm$ 0.08 & 0.93 $\pm$ 0.05 \\
        CSP15B & 2.2 & 13.9 $\pm$ 0.4 & 31.5 $\pm$ 4.5 & 175.4 $\pm$ 5.6 & -19.15 $\pm$ 0.19 & 0.71 $\pm$ 0.05 \\
        CSP15aae & 2.3 & 11.5 $\pm$ 0.4 & 56.8 $\pm$ 2.8 & 151.5 $\pm$ 3.7 & -17.92 $\pm$ 0.15 & 0.47 $\pm$ 0.05 \\
        LSQ11bk & 1.0 & 12.6 $\pm$ 0.3 & 5.6 $\pm$ 1.5 & 98.2 $\pm$ 3.5 & -19.33 $\pm$ 0.06 & 1.08 $\pm$ 0.05 \\
        LSQ12fxd & 3.8 & 10.9 $\pm$ 0.3 & 17.9 $\pm$ 1.9 & 81.0 $\pm$ 3.4 & -19.78 $\pm$ 0.10 & 1.09 $\pm$ 0.05 \\
        LSQ12gdj & 1.6 & 10.7 $\pm$ 0.5 & 8.7 $\pm$ 1.8 & 36.0 $\pm$ 2.2 & -19.78 $\pm$ 0.10 & 1.14 $\pm$ 0.05 \\
        LSQ12hzj & 3.2 & 9.8 $\pm$ 0.3 & 13.5 $\pm$ 2.2 & 59.1 $\pm$ 3.7 & -19.11 $\pm$ 0.10 & 0.96 $\pm$ 0.05 \\
        LSQ13aiz & -6.8 & 14.2 $\pm$ 0.2 & 17.4 $\pm$ 2.7 & 133.6 $\pm$ 3.7 & -19.67 $\pm$ 0.33 & 0.95 $\pm$ 0.05 \\
        LSQ13ry & 4.7 & 12.6 $\pm$ 0.2 & 20.4 $\pm$ 1.2 & 82.3 $\pm$ 1.5 & -19.15 $\pm$ 0.09 & 0.86 $\pm$ 0.05 \\
        LSQ15agh & 3.8 & 9.9 $\pm$ 0.6 & 3.5 $\pm$ 2.6 & 55.3 $\pm$ 5.0 & -19.54 $\pm$ 0.08 & 1.21 $\pm$ 0.05 \\
        LSQ15aja & 3.4 & 10.2 $\pm$ 0.3 & 1.7 $\pm$ 3.2 & 78.1 $\pm$ 8.1 & -19.57 $\pm$ 0.08 & 1.03 $\pm$ 0.05 \\
        PS1-14ra & 3.6 & 11.2 $\pm$ 0.3 & 34.2 $\pm$ 3.9 & 124.0 $\pm$ 6.3 & -19.11 $\pm$ 0.10 & 0.77 $\pm$ 0.05 \\
        PS1-14xw & 1.2 & 11.7 $\pm$ 0.2 & 10.9 $\pm$ 2.1 & 77.8 $\pm$ 2.7 & -19.72 $\pm$ 0.13 & 1.06 $\pm$ 0.05 \\
        PTF11pra & 1.7 & 10.8 $\pm$ 0.4 & 55.5 $\pm$ 3.4 & 126.1 $\pm$ 4.0 & -17.31 $\pm$ 0.24 & 0.40 $\pm$ 0.05 \\
        PTF13duj & 1.6 & 14.2 $\pm$ 0.2 & 7.1 $\pm$ 1.7 & 100.4 $\pm$ 3.4 & -19.35 $\pm$ 0.17 & 1.19 $\pm$ 0.05 \\
        PTF13ebh & 0.6 & 10.8 $\pm$ 0.2 & 48.3 $\pm$ 2.3 & 125.0 $\pm$ 2.7 & -18.76 $\pm$ 0.19 & 0.61 $\pm$ 0.05 \\
        PTF14w & 4.6 & 12.2 $\pm$ 0.3 & 37.0 $\pm$ 2.3 & 124.0 $\pm$ 3.4 & -18.80 $\pm$ 0.13 & 0.73 $\pm$ 0.05 \\
        2004ey & 1.1 & 11.0 $\pm$ 0.3 & 13.8 $\pm$ 4.8 & 100.0 $\pm$ 8.7 & -19.28 $\pm$ 0.17 & 1.01 $\pm$ 0.06 \\
        2004gs & 1.8 & 11.2 $\pm$ 0.1 & 41.7 $\pm$ 6.3 & 136.7 $\pm$ 8.0 & -18.86 $\pm$ 0.11 & 0.69 $\pm$ 0.06 \\
        2005M & 0.5 & 9.2 $\pm$ 0.4 & 7.4 $\pm$ 3.7 & 47.5 $\pm$ 4.5 & -19.61 $\pm$ 0.11 & 1.21 $\pm$ 0.06 \\
        2005bg & 2.6 & 10.6 $\pm$ 0.3 & 7.3 $\pm$ 3.1 & 105.5 $\pm$ 9.9 & -19.47 $\pm$ 0.16 & 1.00 $\pm$ 0.07 \\
        2005el & 1.7 & 10.9 $\pm$ 0.3 & 19.5 $\pm$ 3.2 & 93.9 $\pm$ 4.2 & -19.14 $\pm$ 0.15 & 0.84 $\pm$ 0.06 \\
        2005eq & 6.7 & 9.8 $\pm$ 0.2 & 16.8 $\pm$ 4.8 & 77.3 $\pm$ 7.1 & -19.60 $\pm$ 0.12 & 1.12 $\pm$ 0.06 \\
        2005hc & -6.0 & 10.3 $\pm$ 0.6 & 12.6 $\pm$ 13.5 & 101.5 $\pm$ 16.5 & -19.46 $\pm$ 0.10 & 1.19 $\pm$ 0.06 \\
        2006D & 1.8 & 11.2 $\pm$ 0.2 & 26.4 $\pm$ 0.9 & 100.4 $\pm$ 1.1 & -19.26 $\pm$ 0.24 & 0.81 $\pm$ 0.06 \\
        2006ax & 0.0 & 10.6 $\pm$ 0.3 & 9.6 $\pm$ 2.0 & 92.5 $\pm$ 2.9 & -19.47 $\pm$ 0.13 & 0.99 $\pm$ 0.06 \\
        2006br & 0.6 & 13.9 $\pm$ 0.4 & 8.5 $\pm$ 4.0 & 111.8 $\pm$ 8.5 & -19.52 $\pm$ 0.27 & 0.91 $\pm$ 0.07 \\
        2006hb & 3.2 & 10.5 $\pm$ 0.3 & 41.4 $\pm$ 3.6 & 121.2 $\pm$ 5.2 & -18.83 $\pm$ 0.18 & 0.67 $\pm$ 0.06 \\
        2006hx & 3.5 & 10.5 $\pm$ 0.5 & 12.6 $\pm$ 5.9 & 42.6 $\pm$ 5.6 & -19.64 $\pm$ 0.13 & 0.99 $\pm$ 0.06 \\
        2007S & 0.1 & 10.1 $\pm$ 0.2 & 9.2 $\pm$ 3.2 & 58.7 $\pm$ 5.4 & -19.85 $\pm$ 0.16 & 1.11 $\pm$ 0.06 \\
        2007af & 1.1 & 11.2 $\pm$ 0.2 & 18.7 $\pm$ 5.2 & 104.2 $\pm$ 5.8 & -19.12 $\pm$ 0.35 & 0.93 $\pm$ 0.06 \\
        2007as & 3.5 & 12.6 $\pm$ 0.4 & 18.0 $\pm$ 2.7 & 138.3 $\pm$ 4.7 & -19.24 $\pm$ 0.16 & 0.88 $\pm$ 0.06 \\
        2007ba & 1.7 & 10.8 $\pm$ 0.2 & 49.1 $\pm$ 2.8 & 94.9 $\pm$ 3.0 & -18.66 $\pm$ 0.10 & 0.54 $\pm$ 0.06 \\
        2007bc & 1.1 & 10.6 $\pm$ 0.3 & 29.9 $\pm$ 1.8 & 100.0 $\pm$ 2.4 & -19.32 $\pm$ 0.12 & 0.88 $\pm$ 0.06 \\
        2007bd & 0.3 & 12.7 $\pm$ 0.3 & 12.3 $\pm$ 2.0 & 116.4 $\pm$ 4.4 & -19.28 $\pm$ 0.10 & 0.88 $\pm$ 0.06 \\
        2007bm & 3.6 & 10.6 $\pm$ 0.2 & 24.7 $\pm$ 1.5 & 101.8 $\pm$ 2.7 & -19.59 $\pm$ 0.30 & 0.90 $\pm$ 0.06 \\
        2007ca & 0.9 & 11.0 $\pm$ 0.3 & 10.5 $\pm$ 2.0 & 89.9 $\pm$ 2.4 & -19.56 $\pm$ 0.18 & 1.06 $\pm$ 0.06 \\
        2007le & 2.0 & 11.9 $\pm$ 0.3 & 12.7 $\pm$ 1.6 & 113.9 $\pm$ 3.5 & -19.07 $\pm$ 0.40 & 1.02 $\pm$ 0.06 \\
        2007ol & 2.4 & 11.1 $\pm$ 0.3 & 16.2 $\pm$ 4.1 & 77.2 $\pm$ 8.2 & -18.87 $\pm$ 0.06 & 0.70 $\pm$ 0.07 \\
        2007on & 1.4 & 11.2 $\pm$ 0.1 & 41.3 $\pm$ 3.9 & 117.4 $\pm$ 4.2 & -19.05 $\pm$ 0.35 & 0.57 $\pm$ 0.06 \\
        2007ux & 1.4 & 11.1 $\pm$ 0.2 & 54.3 $\pm$ 7.1 & 124.1 $\pm$ 9.6 & -18.48 $\pm$ 0.10 & 0.59 $\pm$ 0.06 \\
        2008C & 4.3 & 10.6 $\pm$ 0.4 & 15.5 $\pm$ 5.7 & 65.1 $\pm$ 9.2 & -19.38 $\pm$ 0.21 & 0.95 $\pm$ 0.07 \\
        2008O & 0.4 & 14.4 $\pm$ 0.7 & 47.6 $\pm$ 3.4 & 176.3 $\pm$ 4.7 & -18.69 $\pm$ 0.12 & 0.65 $\pm$ 0.06 \\
        2008R & 1.4 & 10.7 $\pm$ 0.2 & 49.4 $\pm$ 1.3 & 126.7 $\pm$ 1.9 & -18.48 $\pm$ 0.18 & 0.59 $\pm$ 0.06 \\
        2008bc & 3.6 & 11.7 $\pm$ 0.2 & 12.6 $\pm$ 3.3 & 105.0 $\pm$ 6.5 & -19.45 $\pm$ 0.16 & 1.05 $\pm$ 0.06 \\
        2008bf & 1.4 & 11.3 $\pm$ 0.2 & 10.7 $\pm$ 2.4 & 79.0 $\pm$ 4.8 & -19.43 $\pm$ 0.10 & 1.02 $\pm$ 0.06 \\
        2008bq & 0.5 & 10.6 $\pm$ 0.2 & 12.4 $\pm$ 3.4 & 92.0 $\pm$ 4.6 & -19.74 $\pm$ 0.16 & 1.16 $\pm$ 0.06 \\
        2008cf & 2.5 & 10.2 $\pm$ 0.3 & 7.9 $\pm$ 2.5 & 55.6 $\pm$ 5.0 & -19.56 $\pm$ 0.10 & 1.12 $\pm$ 0.07 \\
        2008fl & 3.2 & 10.7 $\pm$ 0.2 & 38.3 $\pm$ 4.9 & 112.3 $\pm$ 4.6 & -19.38 $\pm$ 0.18 & 0.85 $\pm$ 0.06 \\
        2008fp & 0.1 & 11.0 $\pm$ 0.1 & 12.4 $\pm$ 3.9 & 73.9 $\pm$ 4.7 & -19.92 $\pm$ 0.39 & 1.08 $\pm$ 0.06 \\
        2008fr & 4.6 & 10.1 $\pm$ 0.4 & 19.3 $\pm$ 3.6 & 93.3 $\pm$ 6.3 & -19.35 $\pm$ 0.12 & 1.06 $\pm$ 0.06 \\
        2008fw & 5.6 & 10.1 $\pm$ 0.2 & 13.0 $\pm$ 3.9 & 61.2 $\pm$ 5.7 & -19.46 $\pm$ 0.27 & 1.11 $\pm$ 0.06 \\
        2008gg & 4.9 & 12.7 $\pm$ 0.4 & 10.0 $\pm$ 3.3 & 149.8 $\pm$ 6.7 & -19.47 $\pm$ 0.15 & 1.11 $\pm$ 0.07 \\
        2008gl & 2.0 & 11.9 $\pm$ 0.4 & 20.6 $\pm$ 2.4 & 115.0 $\pm$ 3.3 & -19.12 $\pm$ 0.09 & 0.85 $\pm$ 0.06 \\
        2008go & 0.1 & 13.1 $\pm$ 0.4 & 8.1 $\pm$ 2.3 & 154.0 $\pm$ 4.9 & -19.32 $\pm$ 0.10 & 0.91 $\pm$ 0.06 \\
        2008gp & 6.8 & 11.0 $\pm$ 0.3 & 15.4 $\pm$ 3.0 & 43.7 $\pm$ 3.1 & -19.42 $\pm$ 0.10 & 0.97 $\pm$ 0.06 \\
        2008hj & 6.4 & 13.0 $\pm$ 0.5 & 16.0 $\pm$ 2.1 & 103.3 $\pm$ 3.5 & -19.30 $\pm$ 0.09 & 1.01 $\pm$ 0.06 \\
        2008hu & 3.7 & 12.4 $\pm$ 0.3 & 15.9 $\pm$ 4.7 & 127.0 $\pm$ 9.4 & -19.04 $\pm$ 0.11 & 0.79 $\pm$ 0.06 \\
        2008hv & 1.3 & 10.8 $\pm$ 0.4 & 21.3 $\pm$ 2.1 & 96.7 $\pm$ 3.3 & -19.12 $\pm$ 0.17 & 0.85 $\pm$ 0.06 \\
        2008ia & 2.3 & 11.4 $\pm$ 0.4 & 20.0 $\pm$ 2.4 & 105.9 $\pm$ 4.2 & -19.31 $\pm$ 0.16 & 0.84 $\pm$ 0.06 \\
        2009D & 0.7 & 9.7 $\pm$ 0.3 & 4.3 $\pm$ 5.3 & 77.5 $\pm$ 6.5 & -19.65 $\pm$ 0.12 & 1.19 $\pm$ 0.06 \\
        2009Y & 2.3 & 14.4 $\pm$ 0.2 & 8.6 $\pm$ 2.0 & 155.6 $\pm$ 4.7 & -19.62 $\pm$ 0.23 & 1.19 $\pm$ 0.06 \\
        2009aa & 0.1 & 10.9 $\pm$ 0.4 & 20.1 $\pm$ 2.9 & 63.2 $\pm$ 2.6 & -19.40 $\pm$ 0.10 & 0.91 $\pm$ 0.06 \\
        2009ab & 2.8 & 10.7 $\pm$ 0.2 & 28.3 $\pm$ 4.1 & 105.2 $\pm$ 5.4 & -19.04 $\pm$ 0.23 & 0.87 $\pm$ 0.06 \\
        2009ad & 1.2 & 10.2 $\pm$ 0.2 & 9.9 $\pm$ 3.8 & 67.4 $\pm$ 3.6 & -19.52 $\pm$ 0.12 & 1.01 $\pm$ 0.06 \\
        2009ag & 1.4 & 10.3 $\pm$ 0.2 & 21.4 $\pm$ 4.5 & 107.8 $\pm$ 5.4 & -19.26 $\pm$ 0.31 & 0.96 $\pm$ 0.06 \\
        2009cz & 0.1 & 10.1 $\pm$ 0.3 & 11.9 $\pm$ 3.2 & 75.6 $\pm$ 4.2 & -19.59 $\pm$ 0.13 & 1.19 $\pm$ 0.06 \\
        2009ds & 4.0 & 12.6 $\pm$ 0.1 & 14.8 $\pm$ 2.3 & 71.8 $\pm$ 3.8 & -19.65 $\pm$ 0.13 & 1.12 $\pm$ 0.06 \\
        2009le & -4.7 & 12.5 $\pm$ 0.2 & 8.2 $\pm$ 1.5 & 79.2 $\pm$ 2.6 & -19.18 $\pm$ 0.16 & 1.16 $\pm$ 0.06 \\
        2011iv & 5.0 & 10.9 $\pm$ 0.2 & 40.8 $\pm$ 2.2 & 88.6 $\pm$ 2.0 & -19.67 $\pm$ 0.34 & 0.64 $\pm$ 0.05 \\
        2011jh & 4.8 & 12.8 $\pm$ 0.2 & 35.0 $\pm$ 2.8 & 126.1 $\pm$ 3.9 & -19.31 $\pm$ 0.29 & 0.80 $\pm$ 0.05 \\
        2012aq & 6.5 & 11.2 $\pm$ 0.4 & 27.6 $\pm$ 2.2 & 105.5 $\pm$ 3.5 & -19.53 $\pm$ 0.15 & 0.99 $\pm$ 0.05 \\
        2012bl & 1.4 & 14.7 $\pm$ 0.3 & 2.3 $\pm$ 1.2 & 90.5 $\pm$ 2.4 & -19.34 $\pm$ 0.14 & 1.11 $\pm$ 0.05 \\
        2012fr & 0.9 & 12.1 $\pm$ 0.2 & 6.5 $\pm$ 1.7 & 70.5 $\pm$ 3.1 & -19.46 $\pm$ 0.40 & 1.12 $\pm$ 0.05 \\
        2012gm & 5.8 & 10.4 $\pm$ 0.3 & 24.1 $\pm$ 4.0 & 101.5 $\pm$ 7.5 & -19.46 $\pm$ 0.17 & 0.98 $\pm$ 0.05 \\
        2012hl & 3.5 & 12.9 $\pm$ 0.2 & 13.3 $\pm$ 3.4 & 142.0 $\pm$ 7.3 & -18.86 $\pm$ 0.27 & 0.92 $\pm$ 0.06 \\
        2012hr & 5.6 & 12.6 $\pm$ 0.3 & 15.3 $\pm$ 1.3 & 134.7 $\pm$ 2.1 & -19.19 $\pm$ 0.29 & 0.96 $\pm$ 0.05 \\
        2012ht & 1.6 & 10.9 $\pm$ 0.2 & 27.5 $\pm$ 1.9 & 115.2 $\pm$ 2.7 & -19.06 $\pm$ 0.61 & 0.85 $\pm$ 0.05 \\
        2012ij & 0.2 & 10.8 $\pm$ 0.3 & 53.3 $\pm$ 4.0 & 121.5 $\pm$ 4.7 & -18.13 $\pm$ 0.21 & 0.53 $\pm$ 0.05 \\
        2013E & 5.9 & 12.6 $\pm$ 0.3 & 3.8 $\pm$ 1.5 & 65.2 $\pm$ 3.0 & -19.90 $\pm$ 0.25 & 1.12 $\pm$ 0.05 \\
        2013fy & 5.1 & 10.8 $\pm$ 0.3 & 14.4 $\pm$ 1.4 & 93.4 $\pm$ 3.1 & -19.67 $\pm$ 0.12 & 1.19 $\pm$ 0.05 \\
        2013gy & 3.1 & 10.2 $\pm$ 0.3 & 28.9 $\pm$ 4.2 & 114.1 $\pm$ 4.3 & -19.39 $\pm$ 0.18 & 0.89 $\pm$ 0.05 \\
        2014I & 1.3 & 11.3 $\pm$ 0.2 & 23.5 $\pm$ 3.7 & 98.7 $\pm$ 4.7 & -19.48 $\pm$ 0.10 & 0.90 $\pm$ 0.05 \\
        2014dn & 3.0 & 10.4 $\pm$ 0.6 & 70.9 $\pm$ 6.2 & 140.1 $\pm$ 4.7 & -17.68 $\pm$ 0.13 & 0.45 $\pm$ 0.05 \\
        \hline
        1998dh & 1.4 & 12.4 $\pm$ 0.2 & 26.1 $\pm$ 2.2 & 124.6 $\pm$ 2.0 & -19.34 $\pm$ 0.22 & \nodata \\
        1998dm & 2.0 & 11.0 $\pm$ 0.2 & 11.2 $\pm$ 1.9 & 73.2 $\pm$ 2.2 & -18.68 $\pm$ 0.28 & \nodata \\
        1999cp & 4.0 & 10.6 $\pm$ 0.2 & 22.7 $\pm$ 1.4 & 104.6 $\pm$ 2.4 & -19.36 $\pm$ 0.18 & \nodata \\
        1999dq & 0.5 & 11.1 $\pm$ 0.1 & 9.3 $\pm$ 1.6 & 44.8 $\pm$ 2.3 & -19.82 $\pm$ 0.15 & \nodata \\
        1999gp & 0.1 & 11.1 $\pm$ 0.4 & 8.2 $\pm$ 2.2 & 53.8 $\pm$ 3.0 & -19.61 $\pm$ 0.09 & \nodata \\
        2000cx & 0.0 & 11.8 $\pm$ 0.3 & 5.9 $\pm$ 2.2 & 40.3 $\pm$ 3.8 & -19.31 $\pm$ 0.24 & \nodata \\
        2000dn & 1.8 & 9.7 $\pm$ 0.3 & 13.3 $\pm$ 3.2 & 102.4 $\pm$ 3.6 & -19.15 $\pm$ 0.07 & \nodata \\
        2000dr & 0.0 & 10.3 $\pm$ 0.3 & 84.6 $\pm$ 3.2 & 131.4 $\pm$ 3.2 & -18.52 $\pm$ 0.11 & \nodata \\
        2000fa & 0.4 & 11.9 $\pm$ 0.3 & 10.7 $\pm$ 2.3 & 83.4 $\pm$ 4.8 & -19.54 $\pm$ 0.10 & \nodata \\
        2001V & 2.5 & 11.4 $\pm$ 0.1 & 13.2 $\pm$ 2.1 & 57.0 $\pm$ 1.9 & -19.70 $\pm$ 0.13 & \nodata \\
        2001en & 0.4 & 12.6 $\pm$ 0.3 & 9.4 $\pm$ 1.5 & 135.0 $\pm$ 3.1 & -18.86 $\pm$ 0.15 & \nodata \\
        2001ep & 0.4 & 10.7 $\pm$ 0.2 & 31.9 $\pm$ 2.1 & 111.7 $\pm$ 2.1 & -19.24 $\pm$ 0.15 & \nodata \\
        2002bo & 2.0 & 13.4 $\pm$ 0.2 & 10.5 $\pm$ 5.1 & 150.4 $\pm$ 9.3 & -19.31 $\pm$ 0.29 & \nodata \\
        2002cr & 0.3 & 10.1 $\pm$ 0.1 & 19.3 $\pm$ 1.8 & 104.6 $\pm$ 2.0 & -19.34 $\pm$ 0.18 & \nodata \\
        2002dj & 5.0 & 14.0 $\pm$ 0.3 & 6.3 $\pm$ 1.3 & 151.0 $\pm$ 2.3 & -19.26 $\pm$ 0.21 & \nodata \\
        2002dl & 6.6 & 12.5 $\pm$ 0.4 & 40.2 $\pm$ 4.5 & 90.1 $\pm$ 4.9 & -18.28 $\pm$ 0.12 & \nodata \\
        2002eb & 3.0 & 10.1 $\pm$ 0.2 & 9.9 $\pm$ 2.1 & 65.0 $\pm$ 4.1 & -19.60 $\pm$ 0.08 & \nodata \\
        2002er & 0.0 & 12.0 $\pm$ 0.0 & 24.2 $\pm$ 0.9 & 115.2 $\pm$ 1.2 & -19.34 $\pm$ 0.21 & \nodata \\
        2002fk & 1.4 & 9.8 $\pm$ 0.3 & 18.6 $\pm$ 1.7 & 80.7 $\pm$ 2.2 & -19.40 $\pm$ 0.24 & \nodata \\
        2002ha & 1.0 & 11.3 $\pm$ 0.3 & 30.6 $\pm$ 2.6 & 111.3 $\pm$ 2.8 & -19.16 $\pm$ 0.14 & \nodata \\
        2002he & 0.6 & 12.6 $\pm$ 0.2 & 20.2 $\pm$ 1.1 & 118.3 $\pm$ 1.5 & -19.01 $\pm$ 0.09 & \nodata \\
        2003W & 0.4 & 15.2 $\pm$ 0.5 & 6.4 $\pm$ 1.7 & 108.2 $\pm$ 3.5 & -19.44 $\pm$ 0.10 & \nodata \\
        2003Y & 2.0 & 11.3 $\pm$ 0.4 & 71.8 $\pm$ 8.2 & 104.1 $\pm$ 5.9 & -17.48 $\pm$ 0.14 & \nodata \\
        2003cg & 0.2 & 11.1 $\pm$ 0.3 & 24.0 $\pm$ 2.5 & 96.1 $\pm$ 3.1 & -19.49 $\pm$ 0.32 & \nodata \\
        2003gn & 3.3 & 13.2 $\pm$ 0.9 & 40.8 $\pm$ 9.3 & 146.8 $\pm$ 10.3 & -18.81 $\pm$ 0.11 & \nodata \\
        2003gt & 4.8 & 11.3 $\pm$ 0.4 & 33.6 $\pm$ 3.6 & 76.8 $\pm$ 4.1 & -19.47 $\pm$ 0.13 & \nodata \\
        2004at & 0.2 & 10.7 $\pm$ 0.2 & 10.5 $\pm$ 1.9 & 92.6 $\pm$ 2.3 & -19.46 $\pm$ 0.09 & \nodata \\
        2004dt & 0.0 & 16.0 $\pm$ 0.0 & 21.5 $\pm$ 1.9 & 167.0 $\pm$ 3.0 & -19.83 $\pm$ 0.10 & \nodata \\
        2005cf & 0.0 & 10.3 $\pm$ 0.3 & 17.0 $\pm$ 1.1 & 89.7 $\pm$ 1.2 & -19.41 $\pm$ 0.25 & \nodata \\
        2005de & 1.7 & 10.7 $\pm$ 0.3 & 30.2 $\pm$ 3.2 & 102.9 $\pm$ 3.5 & -19.06 $\pm$ 0.13 & \nodata \\
        2006cp & 3.7 & 14.9 $\pm$ 0.5 & 15.3 $\pm$ 2.4 & 159.0 $\pm$ 3.0 & -19.27 $\pm$ 0.11 & \nodata \\
        2006gr & 0.3 & 11.3 $\pm$ 0.4 & 3.3 $\pm$ 2.5 & 65.0 $\pm$ 4.2 & -19.41 $\pm$ 0.09 & \nodata \\
        2006le & 2.4 & 11.4 $\pm$ 0.2 & 13.0 $\pm$ 2.4 & 93.3 $\pm$ 5.2 & -19.85 $\pm$ 0.11 & \nodata \\
        2006lf & 2.6 & 11.7 $\pm$ 0.1 & 25.7 $\pm$ 3.0 & 100.4 $\pm$ 4.4 & -19.39 $\pm$ 0.16 & \nodata \\
        2007ci & 0.8 & 12.3 $\pm$ 0.4 & 50.0 $\pm$ 2.6 & 126.4 $\pm$ 2.8 & -18.58 $\pm$ 0.11 & \nodata \\
        2008ec & 0.4 & 10.8 $\pm$ 0.7 & 36.7 $\pm$ 2.8 & 119.3 $\pm$ 3.1 & -19.21 $\pm$ 0.13 & \nodata \\
    \enddata
    \tablecomments{$s_{BV}$ values are not provided for the
                   \citet{Zheng_IaLCIII18} subset and are not used in
                   this work.}
\end{deluxetable*}

\section{Methods}

\subsection{Velocities and Pseudo-Equivalent Widths}

We use a modified version of the \texttt{Spextractor} code
\citep{semeliphd}\footnote{\url{https://github.com/astrobarn/spextractor}}
to measure velocities and pseudo-equivalent widths. We
modified\footnote{\url{https://github.com/anthonyburrow/spextractor}} this
code to allow for downsampling spectral information, with the constraint that
the number of photons is conserved, in order to reduce
computational cost. We have also made adjustments that produce a
more representative Gaussian process regression (GPR) model for a given
spectrum. In the original program, the posterior was sampled at points given to
the prior, whereas now we sample the posterior at uniformly spaced points at a
higher resolution than the prior to account for point-to-point variance. Flux
uncertainties are also added to the Mat\'ern 3/2 GPR kernel in quadrature when
available. The techniques employed are further discussed in
\autoref{sec:appx-GP}.

\begin{figure}[ht]
    \centering
    \includegraphics[scale=1.3]{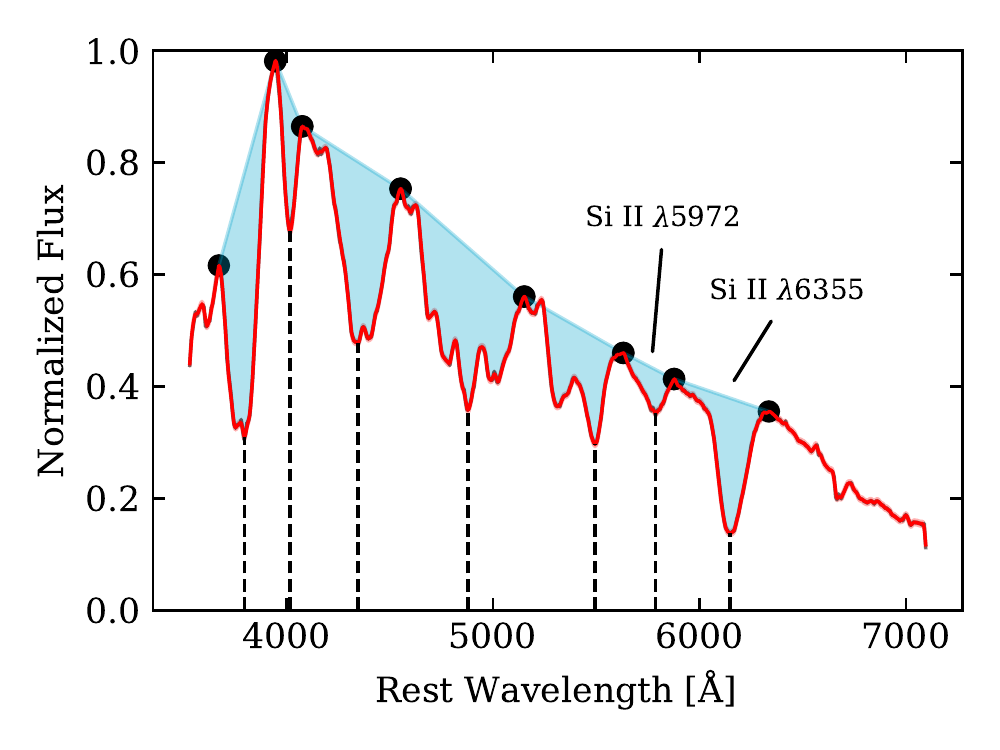}
    \caption{Representative \texttt{Spextractor} function acting on a spectrum
             of ASASSN-14mf at epoch $\sim 5$ days past maximum $B$ light. The
             red line indicates the mean function of the corresponding GPR.
              With a large number of data points, the GPR mean
             function fits well to the original spectrum, which is still
             displayed, albeit difficult to discern from the mean. The blue
             area represents the area integrated in pEW calculations for each
             line. Note that only the \ion{Si}{2} $\lambda5972$ and
             $\lambda6355$ lines are studied in this work, so no special care
             was taken in extracting other features accurately. The
             large black dots indicate the points used to linearly
             determine the pseudo-continuum indicated by the solid blue line.}
    \label{fig:asas14mf}
\end{figure}

\autoref{fig:asas14mf} shows the basic function of the modified
\texttt{Spextractor} code on ASASSN-14mf at epoch $\sim 5$ days past maximum
light. The red curve is the mean function obtained using GPR and the
vertical dashed lines show the position of the flux
minima of identified features. We identify the wavelength of
the flux minima of the
features as the shift due to the (pseudo) photospheric velocities  via
the non-relativistic Doppler formula.
The light blue shading shows the area
used to calculate pseudo-equivalent widths (pEWs). Only specified
features are marked with \texttt{Spextractor}, and in this work
special care is taken only in retrieving accurate measurements for the \ion{Si}{2}
$\lambda5972$ and $\lambda6355$ lines. In order to obtain
velocities we assume that the minima identified correspond with
the rest wavelengths of \Bline and \Rline.  We assume a linear continuum
approximation between maxima of selected wavelength ranges (as indicated by
large black circles in \autoref{fig:asas14mf}) to define features and their pEWs.
The \vsi, pEW(\Bline), and pEW(\Rline) values calculated by
\texttt{Spextractor} for each SNe are given in \autoref{tab:full_data}.

\subsection{Cluster Analysis}

We invoke a distribution-based cluster analysis
using Gaussian
mixture models (GMMs) to support grouping objects with related properties
\citep{day_gmm69,MacPeel00}. These GMMs are expectation-maximization algorithms that
iteratively calculate which input properties maximize the likelihood of a set
of $m$-dimensional ($m$-D) Gaussian distributions fitting a given training set
of data. After this fit is calculated, a total probability distribution is
implicitly given, and so the probability of a point being associated with
each of the $n$ distributions can be determined. This method is
chosen in order
to describe group assignments probabilistically since there appears to be a
continuous distribution of multiple group clusters in our sample of data. There is
no clear way of separating discrete groups with reasonable confidence
 ---  for
example, ensuring that each group is distinct for some $3\sigma$ interval. We
show that by using
Gaussian distributions and, therefore, a well-defined $\sigma$ deviation for
each group, the determined groups overlap within $3\sigma$, and so we would
consider the groups to be connected or non-discrete. The
probability of a SN belonging to any single group
may be comparable to that for another group, even though it is
within $3\sigma$ of either group's mean position.

The assumption is also made here
that each of the measured quantities $M_B$, \vsi, pEW(\Rline), and pEW(\Bline)
can be represented as jointly Gaussian-distributed to first approximation.
This approximation is assumed only to have some measure of similarity between
these four properties. These quantities
are then used as input parameters to the different GMMs calculated and shown in
\autoref{sec:results} which in return yield groupings based on the probability
of an object being associated with each group. The groupings from any GMM
therefore describe similarity between the input quantities from which the GMM
was calculated.

For each of our GMMs, the number of clusters (groups), $n$, assigned has been
determined first with the assumption that each GMM that includes the two pEW
quantities must have at least $n=4$ clusters. Since the original paper
of \citet{branchcomp206} identified four groups, and it has been
customary to break SNe up into the 4 Branch groups, we do not consider
clustering data involving pEW(\Rline) and pEW(\Bline)
with fewer than four groups.

\begin{figure}[ht]
    \centering
    \includegraphics[scale=1.0]{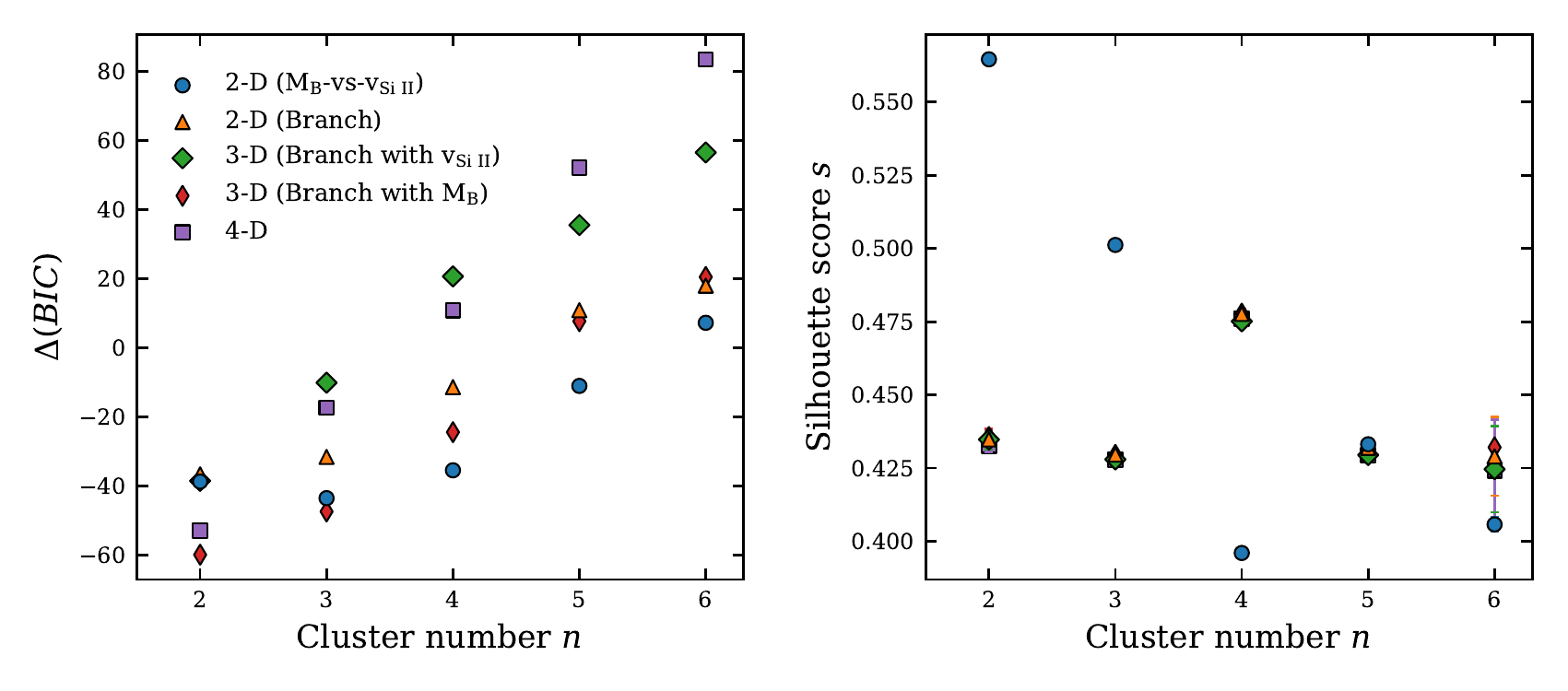}
    \caption{Left panel: $\Delta(\text{BIC})$ versus the cluster number $n$ for each
             GMM presented in this work. $\Delta(\text{BIC})$ is referenced to the
             BIC of the $n=1$ GMM. For all dimensionalities except the [$M_B$,
             \vsi] input set, models with $n < 4$ are not used, in order to
             follow historical precedent \citep{branchcomp206}. Within
             these restrictions, we find
             the $n=4$ case to have the smallest $\Delta(\text{BIC})$,
             thus, these GMMs with $n=4$ are used for this study (see
             \S\S~\ref{sec:branch-clustering}--\ref{sec:higher-dimensional-clustering}).
             For the [$M_B$, \vsi] GMM, the $n=3$ model has the smallest
             $\Delta(\text{BIC})$ value and is therefore used to define
             \polinplot groups discussed in \autoref{sec:polin-plot-comparison}.
             Right panel: the Silhouette score is displayed for each
             model having $n=2$ to $n=6$ clusters.  See text for
             further description. The plot symbols
             and colors are shared between panels.}
    \label{fig:bic}
\end{figure}

In general, the number of clusters used for each GMM may be decided
based on standard testing that determines which value of $n$ provides the model
that best fits the data. For perfectly Gaussian clusters, $n$ may be
determined based on the GMM with the lowest Bayesian Information
Criterion (BIC) value \citep{KR95}. BIC values of each model presented in this
work (see \autoref{sec:results}) were calculated for GMMs with $n=1$ to $n=6$.
Each one of these calculations are the average of 20 single-trial calculations
of the BIC for the model with a given $n$, which is necessary because the GMM
algorithm operates with randomized initial conditions, potentially leading to
different grouping systems for large values of $n$. \autoref{fig:bic}
(left panel) shows
$\Delta(\text{BIC})$ as a function of $n$ for different sets of input
parameters, where $\Delta(\text{BIC})$ references the BIC of the $n=1$
GMM such that $\Delta(\text{BIC}) = \text{BIC}(n) - \text{BIC}(n=1)$.
Therefore, we choose $n$ based on the model yielding the smallest $\Delta(
\text{BIC})$. Results from \autoref{fig:bic} show that all models with
pEW(\Rline) and pEW(\Bline) ("Branch" in the legend, including the 4-D model)
prefer a value of $n=2$. However, because it is assumed that $n \geq 4$ for
these models, we use $n=4$ for these GMMs.

\autoref{fig:bic}
(right panel) shows the mean Silhouette score $s$ \citep{Rousseeuw87,deSouza17}
of the different GMMs for $n=2$ to $n=6$. The values of $s$ were calculated
using $k$-means clustering \citep{macqueen67}. While the Silhouette score is a
good measure of how well the data is separable into clusters, we are
guided by the $\Delta(\text{BIC})$ in determining which value of $n$ to use for
each model. This is because we establish clusters in this sample based on GMMs
and not $k$-means clustering. It is, however, interesting that the Silhouette
score consistently favors $n=4$ ($s$ closer to 1 is more preferable) for any
model involving both pEW(\Rline) and pEW(\Bline).

Neither of these measurements is perfect, and that likely indicates
that the description of maximum-light properties of SNe using Gaussian
distributions is not ideal. However, the Gaussian mixture method's parametric
nature, which allows for a maximum likelihood approach, recommends itself over
other, non-parametric methods. In particular, it provides stable
probabilistic results, which can be interpreted using standard methods \citep{deSouza17}.

The dimensionality, $m$, of each GMM is determined by the
number of SN properties included in training the
GMM, which is independent of the number of GMM components $n$. We
do not weight
any points in the GMMs based on their uncertainty in any quantity.

\section{Results}
\label{sec:results}

\subsection{Branch Clustering}
\label{sec:branch-clustering}

\begin{figure}[ht]
    \centering
    \includegraphics[scale=1.3]{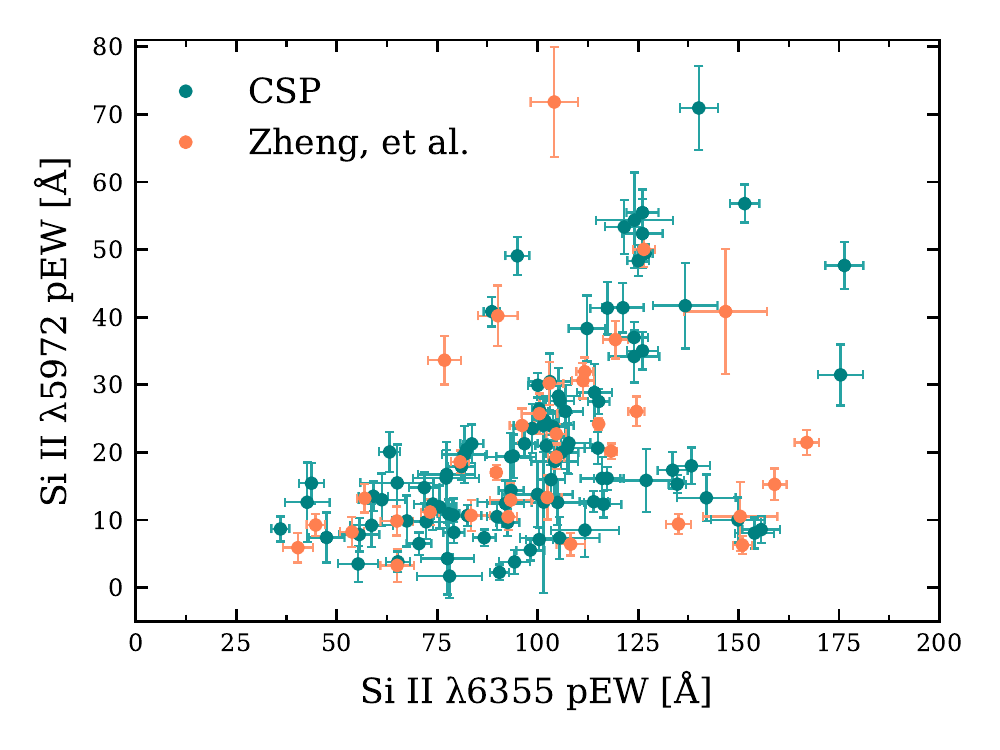}
    \caption{The Branch diagram that includes both the \citet{Zheng_IaLCIII18}
             and the CSP I+II samples. We see the expected Branch diagram trend
             with nine extended cool objects with pEW(\Bline)
             $\geq$ 50 \AA.
             Overall, pEW(\Rline) in this sample exhibits a mean uncertainty of
             $\pm$ 4.35 \AA\ and pEW(\Bline) exhibits that of $\pm$ 3.02
             \AA.}
    \label{fig:p5_p6}
\end{figure}

\autoref{fig:p5_p6} shows the Branch diagram obtained for the CSP I+II and
\citet{Zheng_IaLCIII18} samples. With this large sample of data, it appears
that there are no completely disconnected groupings: a similar and expected
result compared to \citet{branchcomp206}. We do see the expected 0 \AA\
$\leq$ pEW(\Rline) $\leq$ 200 \AA\ range. However, although the majority of
the objects fall in the expected 0 \AA\ $\leq$ pEW(\Bline) $\leq$ 50 \AA\
range, we find  a few extended cools with
pEW(\Bline) $\geq$ 50 \AA. Due to the
lack of a discrete clustering found between these groups, we perform cluster
analysis statistically. Using this entire data set, we create a 2-D GMM in
[pEW(\Bline), pEW(\Rline)] space with (maximal) $n=4$ components,
as it is seen from \autoref{fig:bic} that $n=4$ minimizes
$\Delta(\text{BIC})$ for $n \geq 4$. For this and every other GMM that
includes pEW(\Bline) and pEW(\Rline) in its input parameter space, this is also
the case. The right panel of \autoref{fig:bic} shows that the Silhouette score is maximized for
these models at $n=4$ and
thus supports this number of clusters for this sample. This GMM is
displayed in \autoref{fig:p5_p6_GMM_branch}. Different colors indicate group
membership as a probability distribution of each point belonging to a given
group, and different symbols indicate the group corresponding to highest
likelihood of membership. Contours correspond to 1-, 2-, and $3\sigma$ from
the mean of each group determined by the GMM and is a representation of the
covariance of the GMM groups. This figure clearly identifies four groups that
indeed correspond to the originally identified Branch groups: core-normals
(CN), shallow-silicons (SS), cools (CL), and broad-lines (BL). Note that, in
this and future figures with contours representing group membership, we choose
to exclude error bars for visual purposes, as the same errors are always
displayed in previous figures.

\subsection{Higher-Dimensional GMM Clustering}
\label{sec:higher-dimensional-clustering}

Although the robustness of the Branch groups is seen with a simple 2-D
GMM, in intermediate areas between these groups many objects have comparable
probabilities of membership to more than one group. For this reason, to achieve
more
certainty in membership to any single Branch group, we include additional input
property that are related to the two pEW parameters in the 2-D GMM. This is
expected to provide further constraints for our sample and possibly provide
insight into the Branch groups' relationship with the \polinplot diagram,
which is discussed in \autoref{sec:polin-plot-comparison}.

\subsubsection{Inclusion of \vsi}

\begin{figure}[ht]
    \centering
    \begin{minipage}{\textwidth}
        \centering
        \includegraphics[scale=0.87]{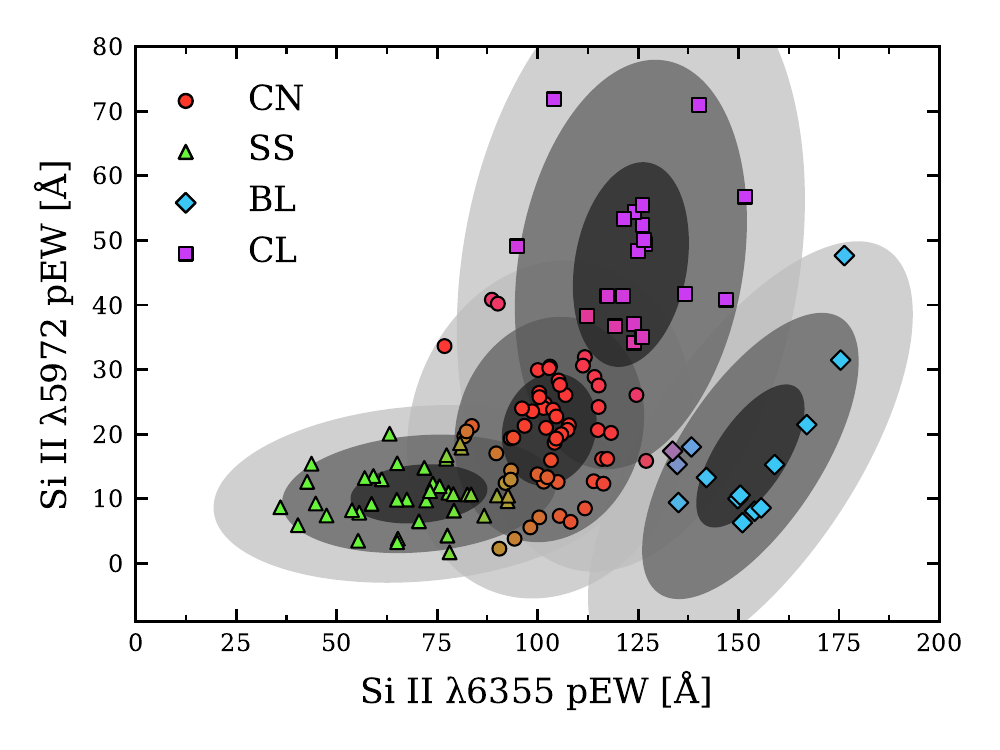}
        \caption{The Branch diagram colored with the [pEW(\Bline), pEW(\Rline)]
                 GMM ($n=4$ components). Contours indicate
                 1-, 2-, and $3\sigma$ from the mean of each group determined
                 by the GMM. Different colors indicate group
membership as a probability distribution of each point belonging to a given
group, and different symbols indicate the group corresponding to highest
likelihood of membership.}
        \label{fig:p5_p6_GMM_branch}
    \end{minipage}
    \begin{minipage}{\textwidth}
        \centering
        \includegraphics[scale=0.98]{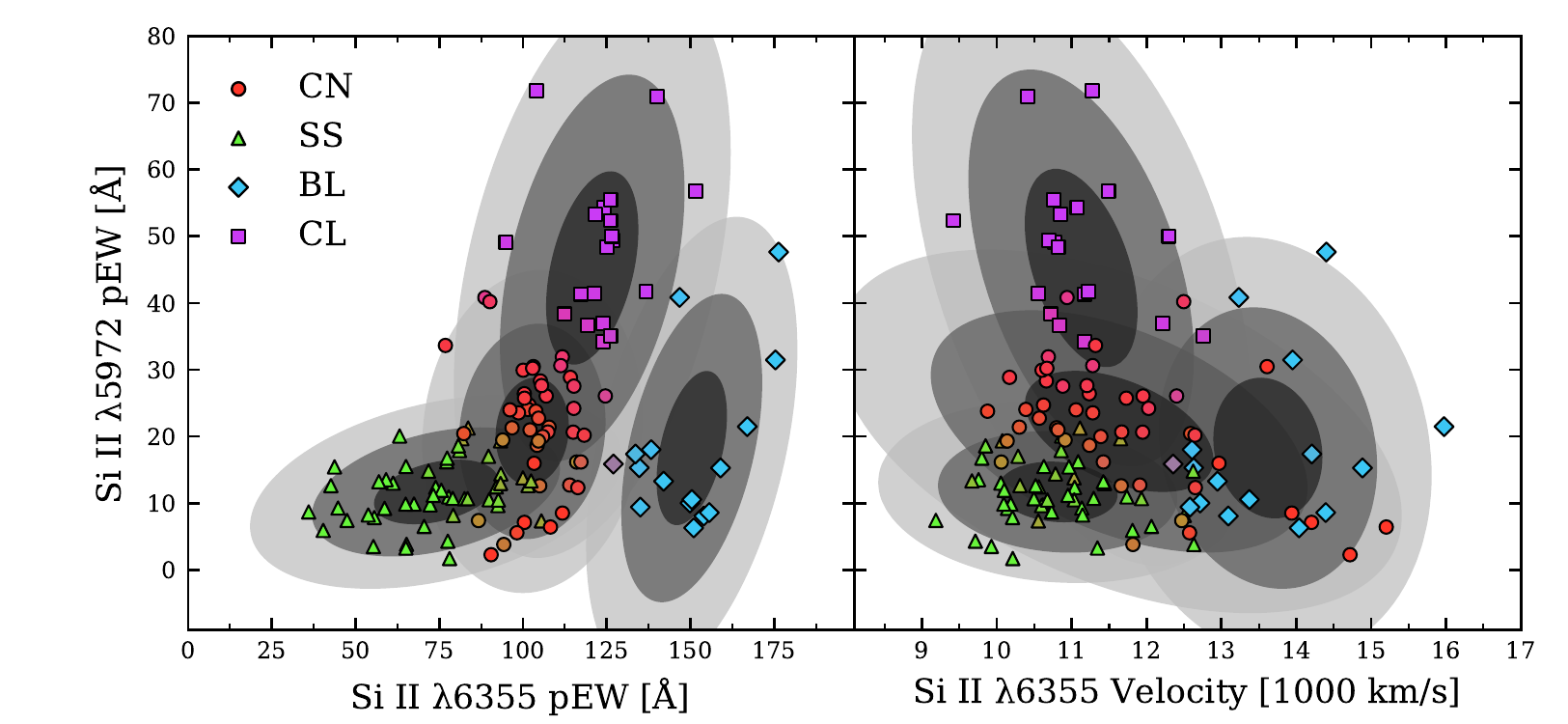}
        \caption{A 3-D GMM analysis of [\vsi, pEW(\Bline), pEW(\Rline)] ($n=4$
                 components). The colors in each panel represent group
                 membership and are shared between panels.}
        \label{fig:v_p5_p6_GMM_v_p5_p6}
    \end{minipage}
    \begin{minipage}{\textwidth}
        \centering
        \includegraphics[scale=0.98]{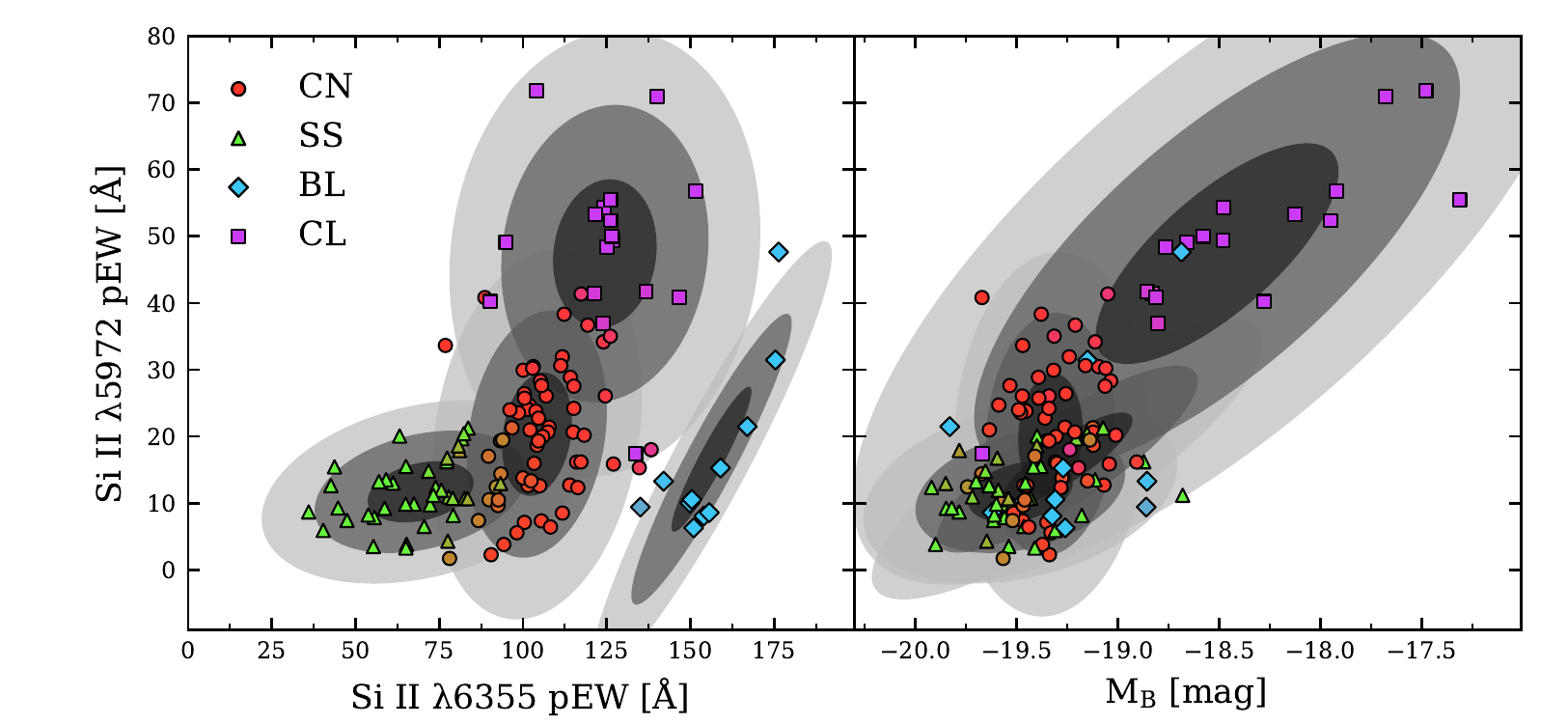}
        \caption{A 3-D GMM analysis of [$M_B$, pEW(\Bline), pEW(\Rline)] ($n=4$
                 components).}
        \label{fig:m_p5_p6_GMM_m_p5_p6}
    \end{minipage}
\end{figure}

We first look at a 3-D GMM in [\vsi, pEW(\Bline), pEW(\Rline)] with $n=4$
components (see \autoref{fig:bic}). \autoref{fig:v_p5_p6_GMM_v_p5_p6}
shows the entire scope of this
model. The 2-D contours in each panel are associated with the covariance of
every determined group distribution for the respective 2-D slice. The left
panel is again a Branch diagram of the sample, and it again clearly separates
the sample into Branch-like groups.

It appears that the inclusion of \vsi information mostly alters the membership
likelihoods in intermediate areas surrounding the CN group. This leads to
changes in the covariances shown by the contours in
\autoref{fig:v_p5_p6_GMM_v_p5_p6} as well as the general probability distribution
(coloring). Comparing with those from the 2-D model in
\autoref{fig:p5_p6_GMM_branch}, we find that the $3\sigma$ contour area of the CN
group decreased by 13.9\%, that of the BL group decreased by 1.8\%, and of the
CL group decreased by 18.1\%, however the $3\sigma$ contour area of the SS
group increased by 8.5\%.

An important difference is that in areas between groups there is a
steeper probability gradient in the 3-D model than the 2-D model in pEW
space. More specifically, we refer to this probability gradient as the gradient
at a given point of the group membership probability distribution projected
into the 2-D subspace, which in this case is the pEW space.
Qualitatively, this effect would narrow
the shape or size of a group's contours, which corresponds to the aforementioned
decreases in the $3\sigma$ areas of the CN, BL, and CL groups. In
intermediate areas between groups it more
concretely defines the group membership of many objects that previously showed
a nearly equal tendency toward two or more groups.
This behavior is expected since naively one would expect a correlation
between the pseudo-equivalent width and velocity of the \Rline line, even
though they are independently measured.

It is therefore seen that using a GMM to measure the similarity between the two
quantities enforces a constraint that quantitatively defines
groups. Most noticeably the contours for the CN and CL groups have reduced in
size, meaning there is a narrower region in pEW space in which CNs are expected
to lie.
This illustrates that including the additional \vsi parameter in the GMM more
sharply defines Branch group membership and leads to more certainty in
assignment compared to a GMM based on only pEW information.

\subsubsection{Inclusion of $M_B$}
\label{sec:3D-MB}

\begin{figure}[ht]
    \centering
    \includegraphics[scale=1.3]{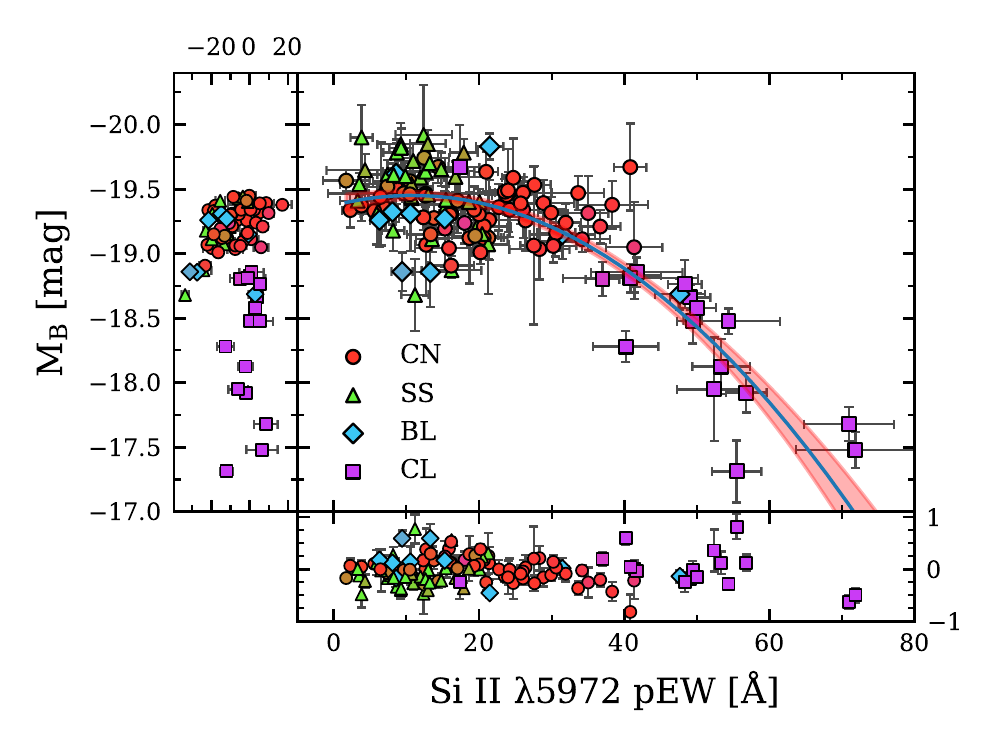}
    \caption{A quadratic fit of the $M_B$ versus pEW(\Bline) relation colored
             with the grouping from the 3-D GMM displayed by
             \autoref{fig:m_p5_p6_GMM_m_p5_p6}. Residuals are given in panels to
             the left and below. The light red shaded region
             represents the uncertainty in the polynomial coefficients
             themselves.  We find the quadratic fit coefficients
             (from highest to lowest order) to be
             $a = (6.623 \pm 1.168) \times 10^{-4}$,
             $b = (-1.425 \pm 0.657) \times 10^{-2}$, and
             $c = -19.37 \pm 0.07$.}
    \label{fig:m_p5_GMM_m_p5_p6}
\end{figure}

We also investigate the 3-D GMM in [$M_B$, pEW(\Bline), pEW(\Rline)] with $n=4$
components (see \autoref{fig:bic}). \autoref{fig:m_p5_p6_GMM_m_p5_p6}
again shows the different slices of this model. In the left panel, we again
find that the Branch diagram shown has
more concretely defined Branch groups than in the 2-D pEW space alone. Again
comparing with the contours in the 2-D model displayed in
\autoref{fig:p5_p6_GMM_branch}, we find that all $3\sigma$ contour areas decrease
in size: CN (7.8\%), SS (10.3\%), BL (59.2\%), and CL (23.3\%). Clearly BL
membership alters more drastically when $M_B$ information is
included in the GMM. The contours of the BLs are much narrower, so membership
is generally contained in a narrower region in pEW space. In fact, with the
inclusion of $M_B$ the BL group becomes an almost completely distinguishable
group, indicating that there may in fact be something distinct about the
progenitor system or explosion mechanism that produces BLs.

It is clear from this GMM that using a Gaussian distribution in this cluster
analysis is not a perfect method in predicting cluster membership. We begin to
see some unexpected behavior in the GMM membership determination. For
example, LSQ13aiz (at pEW(\Rline) $\approx 133$ and pEW(\Bline) $\approx 17$)
does indeed appear too bright to be a cool object as is suggested by the GMM. We
see, then, that variations in one of the input properties can lead to
outlier behavior. As will be seen in \autoref{sec:4D-GMM}, this problem is
partially solved with the inclusion of additional information.

The right panel also shows a strong functional
relation between $M_B$ and pEW(\Bline). We fit a quadratic to this
relation, and this is shown in \autoref{fig:m_p5_GMM_m_p5_p6}. It is interesting
that this correlation, along with the Branch group classification, provides a
rough $M_B$ approximation that is purely based on spectroscopic information.
Others have previously noted that pEW(\Bline) and $\Delta m_{15}$ are
correlated \citep{hach10,folatelli13}, and we find that this effect is
quite robust. Given the relationship between $M_B$ and $\Delta m_{15}(B)$, this
correlation is essentially a spectroscopic variant of the Phillips relation.

\subsubsection{Inclusion of Both \vsi and $M_B$}
\label{sec:4D-GMM}

\begin{figure}[ht]
    \centering
    \includegraphics[scale=0.8]{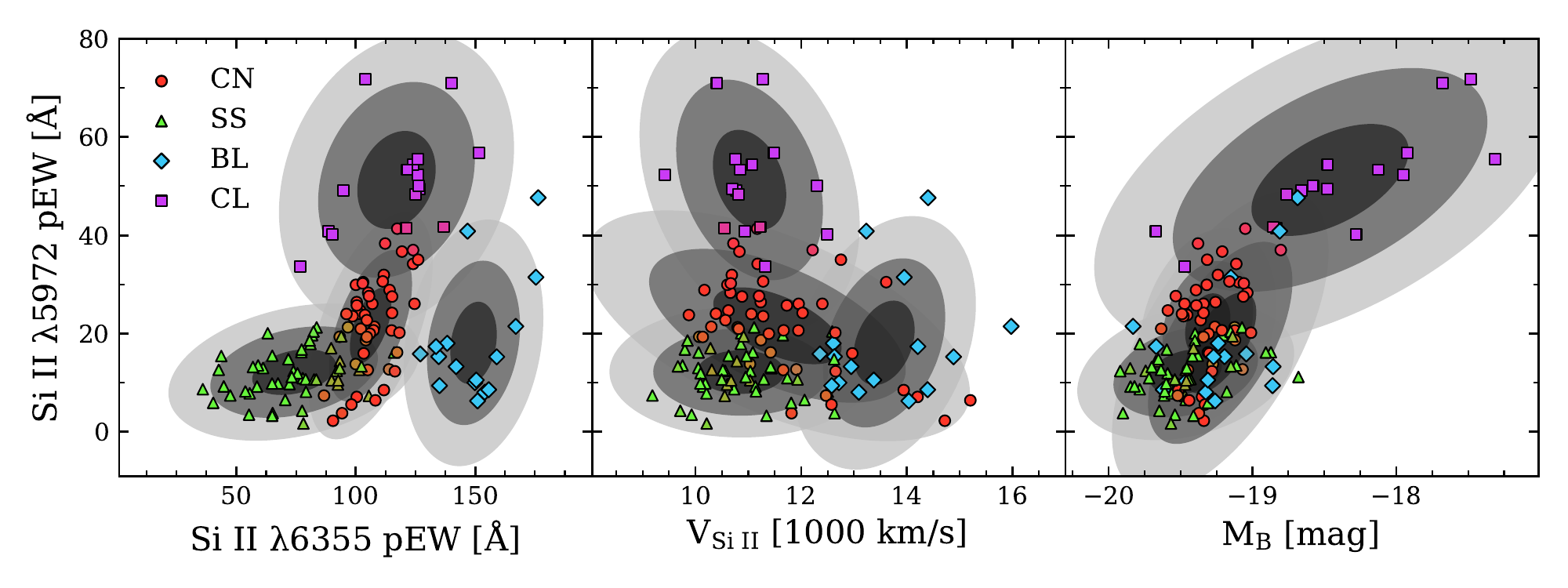}
    \caption{The Branch diagram (left) of the 4-D GMM analysis ($n=4$
             components) of all four parameters, showing the robustness of the
             Branch grouping system. The pEW(\Bline)-versus-\vsi (middle) and
             pEW(\Bline)-versus-$M_B$ (right) projections are shown and are
             color-coded by the same 4-D GMM group membership probabilities.
             The contours indicate 1-, 2-, and $3\sigma$ of group
             membership projected into each respective space. We see the
             groups have become much more constrained and display much overlap
             that could not be extracted solely with pEW information.}
    \label{fig:m_v_p5_p6_GMM_m_v_p5_p6}
\end{figure}

Because we see a strong correlation in [$M_B$, pEW(\Bline)], we attempt to
constrain the Branch groups further by creating a 4-D GMM in [$M_B$, \vsi,
pEW(\Bline), pEW(\Rline)] with $n=4$ components (see
\autoref{fig:bic}). We show this Branch diagram in
the left panel of \autoref{fig:m_v_p5_p6_GMM_m_v_p5_p6}, which again is colored by
this 4-D GMM. This GMM indeed still produces the four Branch
groups. Even with all four dimensions there is still significant overlap
between the CN and SS objects.

The inclusion of both $M_B$ and \vsi constrains the groups further by similarly
reducing the size of most contours associated with projected covariance, giving
a more concrete assignment to more objects. This overall inclusion decreases
the $3\sigma$ contour areas of the CN group (45.6\%), the BL group (28.9\%),
and the CL group (28.7\%). However, the $3\sigma$ contour area of the SS group
is relatively unchanged, increasing by only 0.6\%. In the
projection
shown, there is more overlap
between groups than either 3-D model, which shows that, for our sample, $M_B$
and \vsi contain independent information. That is to say, after including
either quantity, including the fourth quantity will constrain the groups
further.

Again we see that there are some objects that exhibit relatively substantial
dispersion in one or more parameters that appear as outliers. For
example, in the middle panel of \autoref{fig:m_v_p5_p6_GMM_m_v_p5_p6}, SN~2008O
(with \vsi $\sim 14.4$ and pEW(\Bline) $\sim 47.6$) and SN~2003gn (with
\vsi $\sim 13.2$ and pEW(\Bline) $\sim 40.8$) seem most similar to the CL
group, however they are deemed BL objects due to their pEW(\Rline) and \vsi
values.

Compared to the [pEW(\Bline), pEW(\Rline)] GMM defined in
\autoref{fig:p5_p6_GMM_branch}, the inclusion of $M_B$ and \vsi in the GMM
allows us to constrain group membership in a way that is not apparent with only
pEW(\Bline) and pEW(\Rline). That is to say, $M_B$ and \vsi may be used to more
concretely define Branch groups \citep[remembering that we have a completely
different definition for membership that was used by][]{branchcomp206}, between
which there would otherwise be more uncertainty and continuity in the
probability distribution in pEW space.

\subsection{\polinplot Clustering}
\label{sec:polin-plot-comparison}

\begin{figure}[ht]
    \centering
    \includegraphics[scale=1.3]{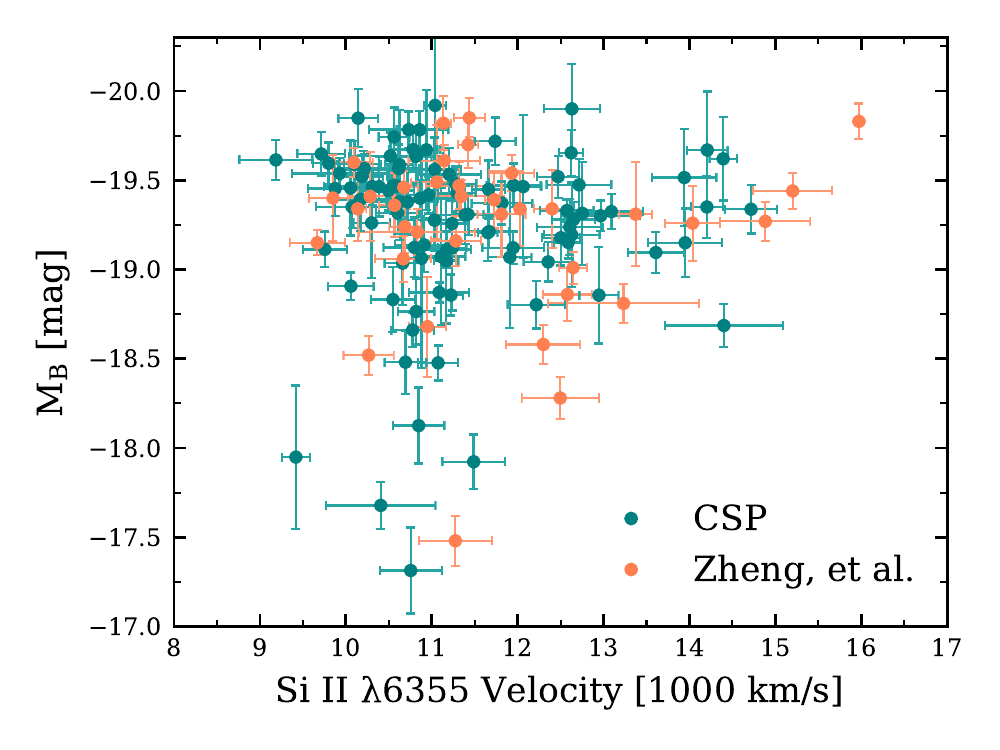}
    \caption{\polinplot diagram of both \citet{Zheng_IaLCIII18} and CSP I+II
             samples. As the CSP sample is included, there are no longer two
             distinct groups, but rather a continuity of a similar form to the
             original \polinplot diagram. For this sample we find $M_B$ to have
             a mean uncertainty of $\pm$ 0.169 mag and \vsi to have that of
             $\pm$ 290 km s$^{-1}$.}
    \label{fig:m_v}
\end{figure}

\begin{figure}[ht]
    \centering
    \includegraphics[scale=1.3]{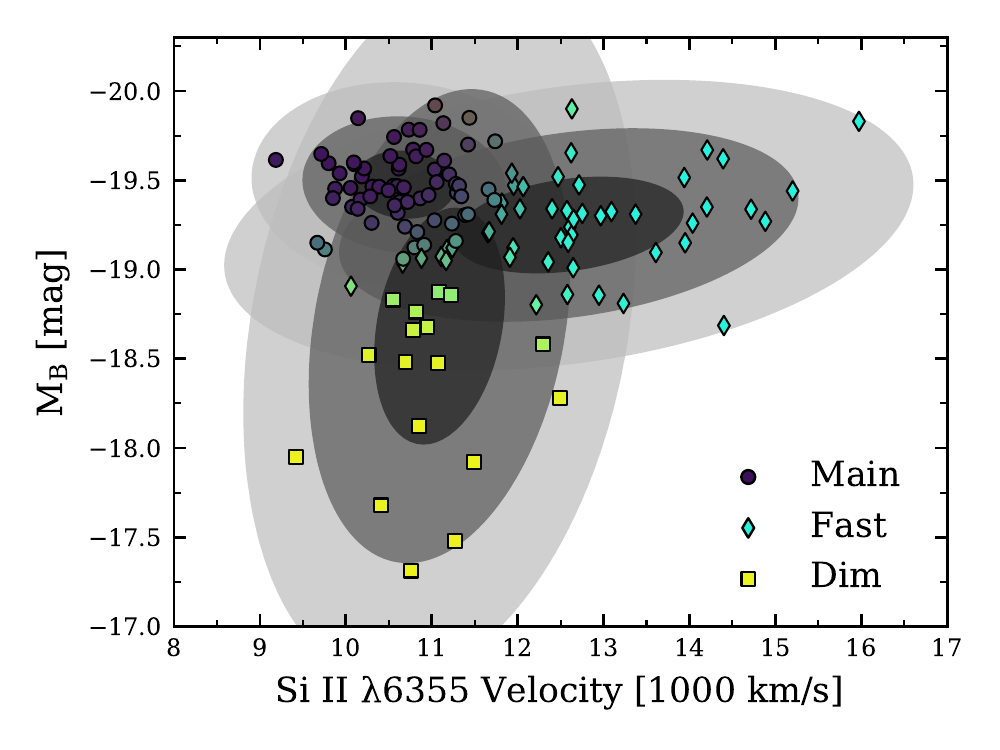}
    \caption{The \polinplot diagram colored by a 2-D GMM in [$M_B$, \vsi]
             ($n=3$ components) which separates the sample into what we label
             as the Main, Dim, and Fast groups. The contours
             correspond to covariance of group membership up to $3\sigma$.}
    \label{fig:m_v_GMM_polin}
\end{figure}

\begin{figure}[ht]
    \centering
    \includegraphics[scale=1.3]{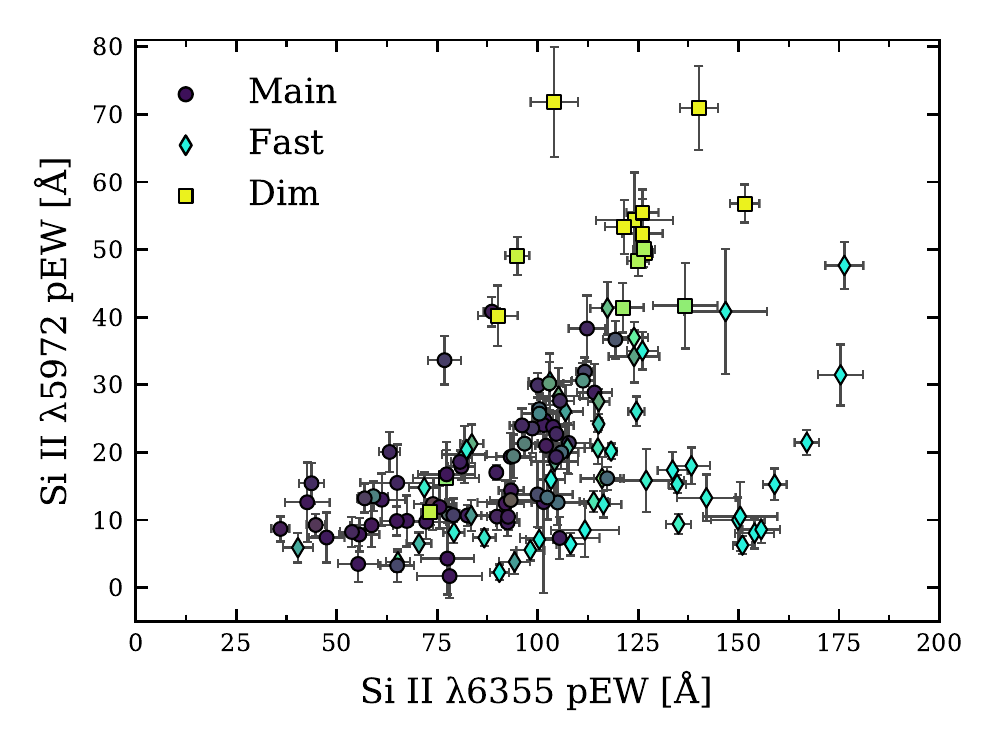}
    \caption{The Branch diagram with groups defined by the 2-D GMM displayed in
             \autoref{fig:m_v_GMM_polin}. We see that there is generally no
             absolute match between the original \citet{polin19hedet} groups and the Branch
             groups as we see many Fast objects contained in BLs as well as
             CNs and SSs.}
    \label{fig:p5_p6_GMM_polin}
\end{figure}

Our version of the \polinplot diagram with both \citet{Zheng_IaLCIII18} and CSP
I+II samples is shown in \autoref{fig:m_v}. The general structure
of the original plot remains, although the CSP sample extends and
fills in the
low- to mid-velocity and dim portions of the plot. We see that the added
continuity removes the original notion of a clear dichotomy between likely
Chandrasekhar mass explosions and sub-Chandrasekhar mass helium detonations
\citep{polin19hedet}. We use a 2-D GMM in [$M_B$, \vsi] to account for this and
to describe the \polinplot group analysis statistically.

\autoref{fig:m_v_GMM_polin} shows the \polinplot diagram with this GMM with $n=3$
components. This value of $n=3$ minimizes $\Delta(\text{BIC})$
(see \autoref{fig:bic}). While the Silhouette score favors two groups for the
\polinplot space, we choose this number based on $\Delta(\text{BIC})$, which
was calculated using a GMM instead of $k$-means clustering. Again, the lack of
a strong preference is likely an indication of the inadequacy of the
description of the groups with Gaussian distributions. We refer to these three
determined groups as the \polinplot groups --- namely the Main group, the Dim
group, and the Fast group.

The Main \polinplot group seems to resemble very closely the more populous
group of SNe in the original \polinplot diagram that \citet{polin19hedet}
interpreted as near-Chandrasekhar-mass explosions. Conversely, the Dim and Fast
groups together make-up the entirety of objects that \citet{polin19hedet}
identified with sub-Chandrasekhar models produced with a thin helium shell.

One may naively expect the Main group to have good overlap with the CN
Branch group, the Fast group to overlap with the BLs [as with high \vsi one
would expect large pEW(\Rline)], and the Dim group to overlap with the CLs.
\autoref{fig:p5_p6_GMM_polin} shows the Branch diagram colored with the \polinplot
groups defined by the GMM illustrated in \autoref{fig:m_v_GMM_polin}. Comparing
with our 4-D GMM description of the Branch groups in
\autoref{fig:m_v_p5_p6_GMM_m_v_p5_p6}, it is clear that there is not a good match
between the \polinplot groups defined by [$M_B$, \vsi] and the Branch groups.
The Main group is made up mostly of both CN and SS SNe. However, we also see
many Fast group objects are either CNs or SSs. This is actually quite
surprising, since \emph{a priori} one would expect a strict relationship
between the high-velocity \polinplot group and the BLs. We see, then,
that there is much dispersion in the relationship between \vsi and pEW(\Rline).
The inconsistency here must be that the intermediate area between the Main and
Fast groups cannot be established exclusively in [$M_B$, \vsi]. Finally, we
remark that the Dim group tends to associate with the CLs nearly entirely,
which is expected.

\begin{figure}[ht]
    \centering
    \includegraphics[scale=0.9]{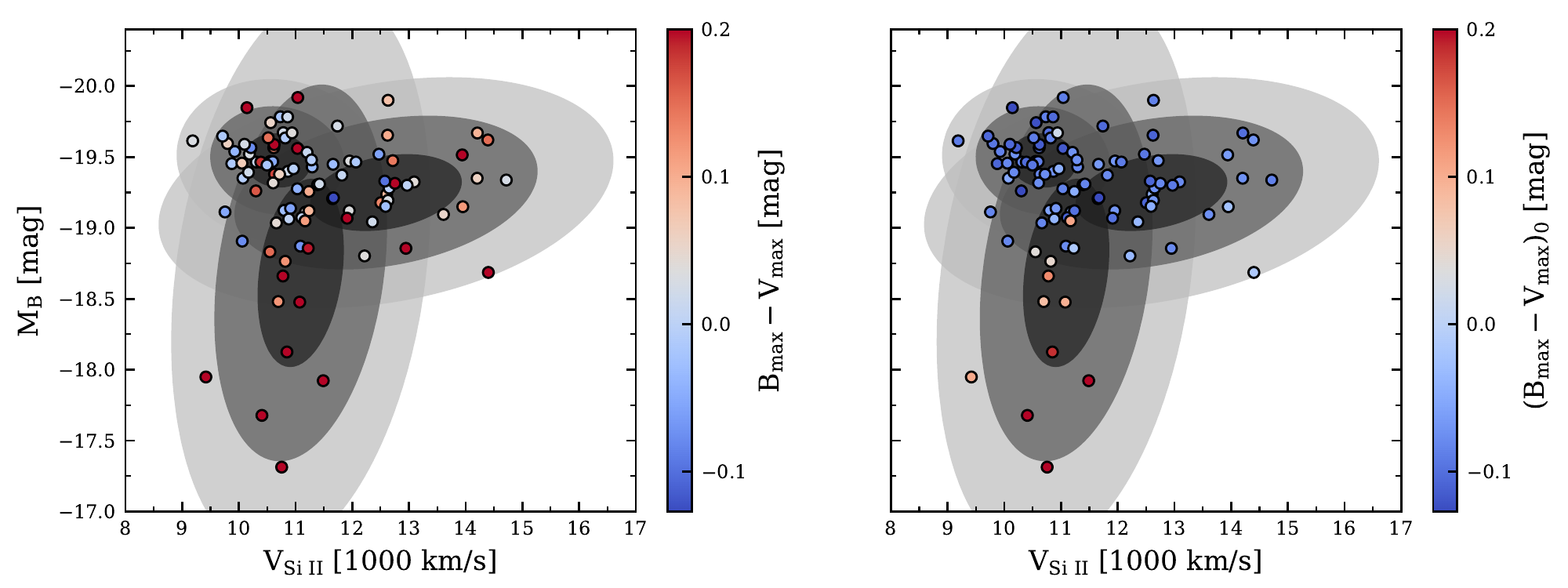}
    \caption{\polinplot diagram of the CSP I+II samples (without the
             \citet{Zheng_IaLCIII18} sample, as they did
             not provide colors for their sample). The left panel is coded for
             $B_\text{max} - V_\text{max}$ color that is uncorrected for host
             galaxy extinction, whereas the right panel is coded for \BV color
             that is corrected for host extinction. The color scale is
             truncated for \BV $> 0.2~\text{mag}$ due to relatively large separation of 19
             outliers with $0.2~\text{mag} < B_\text{max} - V_\text{max}
             \leq 0.939~\text{mag}$ (left panel) and four outliers with $0.2~
             \text{mag} <$ \BV $\leq 0.536$~mag (right panel). The contours are
             exactly those from \autoref{fig:m_v_GMM_polin}, which indicate the
             [$M_B$, \vsi] GMM covariance. $M_B$ plotted in both
             panels is the same, always including host extinction.}
    \label{fig:m_v_BV_GMM_polin_COMP}
\end{figure}

\begin{figure}[ht]
    \centering
    \includegraphics[scale=1.1]{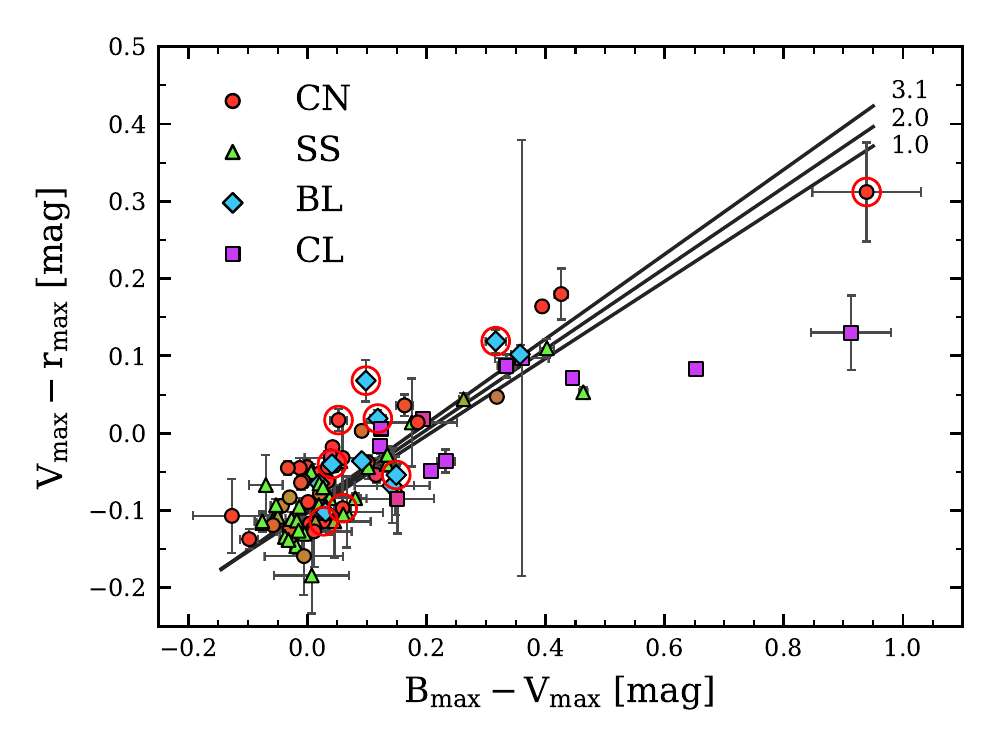}
    \caption{Color-color figure of MW extinction-corrected $V_\text{max}-
             r_\text{max}$ versus
             $B_\text{max}-V_\text{max}$ for the CSP I+II subsample. The figure
             is colored by 4-D Branch group membership, and red circles
             indicate objects with \vsi $>$ 13,000 km s$^{-1}$. $R_V$ values of
             1.0, 2.0, and 3.1 are indicated. The $R_V$ lines show the
             general direction of displacement due to a correction for
             reddening. In general, non-CL SNe are
             generally insensitive to the value of $R_V$ and follow a
             trajectory expected from a dust reddening law. These Fast group
             SNe also follow the dust trajectory, suggesting that for some
             reason they tend to be relatively highly extinguished in their
             host galaxies. While this does not prove that they
             have the correct $M_B$ determined by SNooPy, it makes it plausible.}
    \label{fig:color_color}
\end{figure}

\begin{figure}[ht]
    \centering
    \includegraphics[scale=1.3]{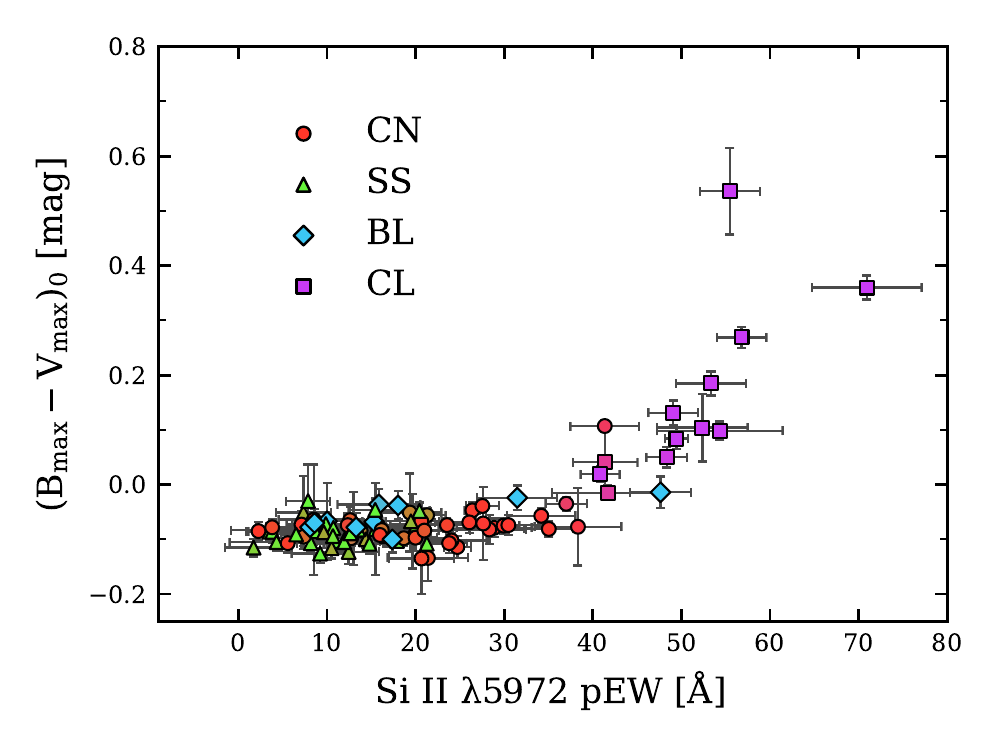}
    \caption{\BV versus pEW(\Bline) of the CSP I+II samples with 4-D Branch
             groups colored. We find that at pEW(\Bline) $\lesssim$ 40 \AA,
             there is no correlation between the two quantities, which is
             expected by the $M_B$ versus pEW(\Bline) relation we find in
             \autoref{sec:3D-MB}.}
    \label{fig:BV_p5_GMM_m_v_p5_p6}
\end{figure}

\begin{figure}[ht]
    \centering
    \includegraphics[scale=1.3]{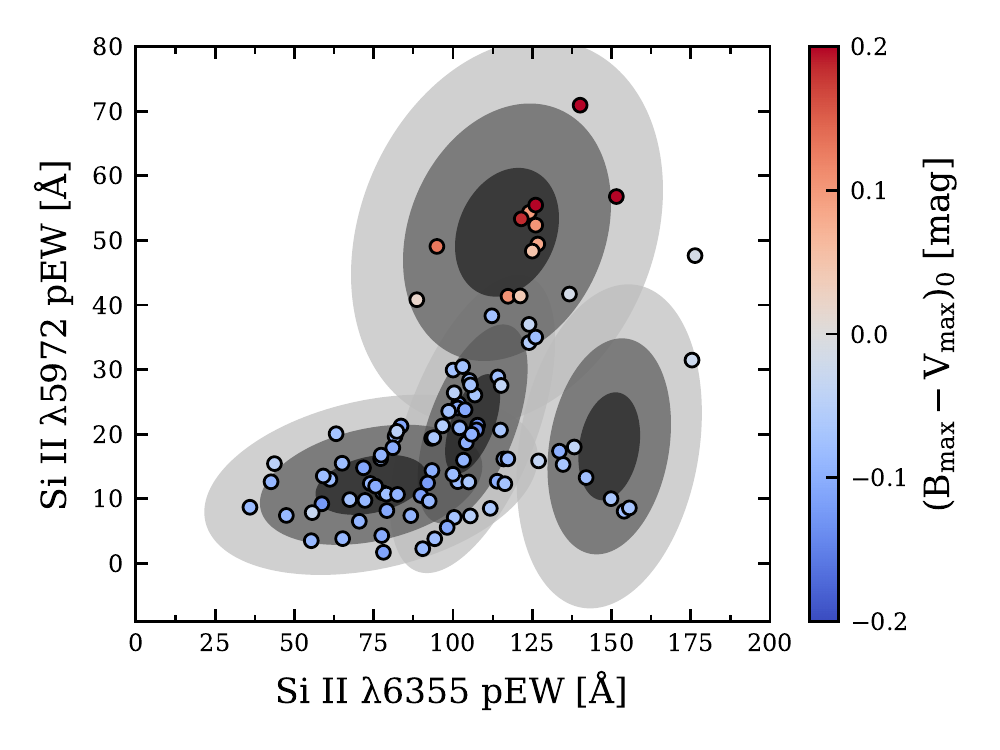}
    \caption{Branch diagram of the CSP I+II samples coded for \BV color. The
             color scale is truncated for \BV $> 0.2$ in the same way as
             \autoref{fig:m_v_BV_GMM_polin_COMP}. The contours are exactly those
             from \autoref{fig:m_v_p5_p6_GMM_m_v_p5_p6}, indicating our defined
             4-D Branch groups.}
    \label{fig:p5_p6_BV_GMM_m_v_p5_p6}
\end{figure}

In \autoref{fig:m_v_BV_GMM_polin_COMP} we show \polinplot diagrams for CSP I+II
data, excluding the \citet{Zheng_IaLCIII18} sample, as colors were not
provided for this sample. The left panel is coded for $B_\text{max}-
V_\text{max}$ color that is uncorrected for host galaxy extinction, and the
right panel is coded for \BV which is corrected for host extinction
\citep{Burns_CSP18}. Both colors are corrected for MW extinction.
Contours in this figure are identical to those found in the [$M_B$, \vsi] GMM
from \autoref{fig:m_v_GMM_polin}. From the left panel we see the similar result
from \citet{polin19hedet} that the bluest objects are consistently found within
the Main group. However, comparing the two, we see that, after correcting for
both MW and host extinction, \BV only seems to have a dependence on $M_B$.
We find no evidence of red objects which are intrinsically bright. It is evident
that both the Main and Fast groups primarily contain intrinsically bluer
objects than the Dim group. After correcting for host extinction, the group
that was claimed to follow sub-Chandrasekhar-mass models by
\citet{polin19hedet} does not display a more consistent redder color than the
more populated clump of Main group objects, indicating that these high velocity
objects are not likely to come from sub-Chandrasekhar-mass explosions.

\subsubsection{Color of the Fast Group SNe}

\begin{deluxetable}{llll}
    \tablecaption{Subset of Fast CSP I+II SNe with \vsi $>$ 13,000 km s$^{-1}$.
        \label{tab:fast}}
    \tablewidth{0pt}
    \tablehead{
        \colhead{SN} & \colhead{\BV} & \colhead{$A_{V}^\text{MW}$}
            & \colhead{$A_{V}^\text{Host}$} \\
        \colhead{} & \colhead{(mag)} & \colhead{(mag)} & \colhead{(mag)}
    }
    \decimals
    \startdata
        ASASSN-14hr & $-0.074 \pm 0.018$ & 0.04 & 0.31 \\
        CSP15B & $-0.024 \pm 0.022$ & 0.20 & 0.36 \\
        LSQ13aiz & $-0.100 \pm 0.024$ & 0.24 & 0.39 \\
        PTF13duj & $-0.073 \pm 0.024$ & 0.21 & 0.15 \\
        2006br & $-0.064 \pm 0.101$ & 0.06 & 2.28 \\
        2008go & $-0.078 \pm 0.016$ & 0.10 & 0.09 \\
        2008O & $-0.014 \pm 0.029$ & 0.24 & 0.37 \\
        2009Y & $-0.070 \pm 0.025$ & 0.27 & 0.28 \\
        2012bl & $-0.085 \pm 0.017$ & 0.09 & 0.01 \\
    \enddata
    \end{deluxetable}

It is seen from \autoref{fig:m_v_BV_GMM_polin_COMP} that nine Fast
objects (chosen by \vsi $>$ 13,000 km s$^{-1}$) from the CSP I+II sample generally
exhibit a redder $B_\text{max}-V_\text{max}$ color (when not corrected
for host galaxy extinction). These SNe are listed in \autoref{tab:fast}, along
with their \BV color and extinction values for Milky Way and host galaxies.
This is in line with the results from \citet{polin19hedet} who showed a similar
plot to the left panel of \autoref{fig:m_v_BV_GMM_polin_COMP}. Note that
\citet{polin19hedet} do not correct for host galaxy extinction. Comparing the
two panels of \autoref{fig:m_v_BV_GMM_polin_COMP} suggests, then, that these Fast
objects seem to be quite reddened in their host galaxy. \autoref{fig:color_color}
shows that in [$V_\text{max} - r_\text{max}$, $B_\text{max}-V_\text{max}$]
space, non-CL SNe generally follow a line that is relatively insensitive to the
value of $R_V$, giving confidence that these Fast SNe are for some reason
preferentially reddened in their host galaxies and that the inferred host
extinction is not due to some underlying assumption in the SNooPy templates
that is misinterpreted. Since \citet{Zheng_IaLCIII18} used MLCSk2 and a fixed
$R_V$ to infer host extinction and $M_B$, although similar but not
identical to the methods of SNooPy, it lends some
credence to the argument that SNooPy is not somehow mis-correcting the Fast
subsample. However, we have introduced a correlation between
pEW(\Bline) and $M_B$ as well as pEW(\Bline) and \BV just by using
SNooPy to fit for these quantities. The same would be true if we were
to use MLCS2k2 or SALT, because their light-curve shape parameters would
also be highly correlated with pEW(\Bline). Nonetheless while these objects
have distinct spectroscopic
properties, we are assuming they all follow the same intrinsic
color--$s_{BV}$ relation from \citet{Burns_CSP14} in order to perform the
extinction corrections.

\citet{xwang09} noted this importance of \vsi, splitting SNe into two groups
and noting that the group with \vsi $\gtrsim$ 11,800 km s$^{–1}$ preferred a
lower value of $R_V \sim 1.6$. \autoref{fig:color_color} shows that these Fast SNe
follow a dust trajectory which is insensitive to $R_V$, which is an indication
that the trend seen in \autoref{fig:m_v_BV_GMM_polin_COMP} is likely due to
extinction from dust in the host. This preference for these Fast objects to
appear in dusty regions \citep[perhaps near the core, see][]{shuvocsppaper1}
should be the subject of future work.

We also note that in contrast to the $B_\text{max}-V_\text{max}$ values
presented in \citet{polin19hedet}, when corrected for host extinction the \BV
range is much narrower, mostly falling within $-0.2 <$ \BV $< 0.2$, having only
four outliers with \BV $> 0.2$ (shown more explicitly in
\autoref{fig:BV_p5_GMM_m_v_p5_p6}). \autoref{fig:BV_p5_GMM_m_v_p5_p6} shows \BV
versus pEW(\Bline) for the CSP I+II sample. We see that at low pEW(\Bline)
there is no correlation between the two quantities. This is expected from
\autoref{fig:m_v_BV_GMM_polin_COMP}, because only Dim objects have a large
increase in \BV compared to brighter objects from the Main and Fast groups that
exhibit bluer color with no other noticeable trend. Generally there is only a
spread of about 0.15 magnitude in \BV for most SNe~Ia.

We also include a Branch diagram colored in \BV for reference to the Branch
groups in \autoref{fig:p5_p6_BV_GMM_m_v_p5_p6}. Contours in this figure are
exactly those determined by the 4-D GMM that are displayed in
\autoref{fig:m_v_p5_p6_GMM_m_v_p5_p6}. We see that, as expected, the CLs are
generally the reddest objects, where there is no other trend between the other
three groups.

\section{Discussion}

\subsection{Constraining the Branch GMM}

Supplementing the two quantities pEW(\Bline) and pEW(\Rline) with more
information  provides more certainty in Branch group assignment.
Quantitatively, the 4-D GMM has been shown to decrease the size of GMM covariance
contours by making a constraint that values must also be similar in additional
dimensions. When we compare the three different models in
\autoref{sec:higher-dimensional-clustering}, however, we see that uncertainty
in the [pEW(\Bline), pEW(\Rline)] GMM mostly decreases with the inclusion of
only a single quantity. Therefore, although we take the 4-D GMM as our defining
model for Branch groups as it includes additional $M_B$ information that we
show is non-linearly correlated with pEW(\Bline), it can still be approximated
with only the inclusion of \vsi. In this way, the Branch groups can be
approximated purely with spectroscopic data. The effects obtained above are
interesting given works showing that the inclusion of \vsi reduces the scatter
of Hubble residuals \citep{foley-kasen11} and that there appears to be a
correlation in the sign of the Hubble residuals and the value of \vsi
\citep{siebert20}. Recently, it was suggested that SNe~Ia with high \vsi are
closely associated with massive host environments and that high \vsi is due to
a variation in explosion mechanism \citep{pan_sn1ahighvel20}. We find that the
BL group is more likely to be distinct, but it would be good to identify the
environments of objects in different Branch groups. We do note that the
tendency of the Fast objects to be strongly extinguished in their
hosts is likely an indication about the nature of the environment of
these SNe~Ia.

It is also seen in \autoref{fig:m_v_p5_p6_GMM_m_v_p5_p6} that 26 SNe change their
most probable group assignment between the 2-D GMM, the 3-D GMM with \vsi
inclusion, and the 4-D GMM. \autoref{tab:changes} details the changes involved
for this sample. Out of the total 26 SNe, 24 are reassigned with the inclusion
of both $M_B$ and \vsi to the GMM, which is substantial compared to the total
sample size of 133 SNe. We find that including these two parameters has the
greatest effect on CN and CL objects, with little to no effect on SS and BL
objects that are defined in the 2-D model. This may be due to the relationship
we find between $M_B$ and pEW(\Bline) coupled with a dispersion in the
\vsi--pEW(\Rline) relation, however more work must be done to
determine the dominant quantities that govern the changes between each group.
Ultimately, group assignment and any physical
attributes that objects of each group may possess are not solely a function of
pseudo-equivalent widths, but of also their other properties such as $M_B$ and
\vsi, as is shown in this work. Interestingly, the classification
changes from CN to SS are fairly stable going from the 3-D GMM to the
4-D GMM, and also the identification of BL is stable from 3-D to
4-D. The most noticeable difference from 3-D to 4-D is CL $\to$ CN for a half-dozen
supernovae.

\begin{deluxetable}{llll}
    \tablecaption{SNe with Branch group membership changes between the 2-D GMM,
                  the spectroscopic 3-D GMM with \vsi dependence, and the 4-D
                  GMM. The group for each model is based on the most probable
                  assignment determined by each GMM.\label{tab:changes}}
    \tablewidth{0pt}
    \tablehead{
        \colhead{SN} & \colhead{2-D} & \colhead{3-D (\vsi)} & \colhead{4-D} \\
        \colhead{} & \colhead{Group} & \colhead{Group} & \colhead{Group}
    }
    \decimals
    \startdata
        CSP14acl & CN & CN & SS \\
        LSQ13ry & CN & CN & SS \\
        2005el & CN & CN & SS \\
        2004ey & CN & SS & CN \\
        2008fr & CN & SS & CN \\
        ASASSN-15al & CN & SS & SS \\
        ASASSN-15hf & CN & SS & SS \\
        2000dn & CN & SS & SS \\
        2005bg & CN & SS & SS \\
        2005cf & CN & SS & SS \\
        2005hc & CN & SS & SS \\
        2006le & CN & SS & SS \\
        2008bq & CN & SS & SS \\
        2013fy & CN & SS & SS \\
        2008hu & CN & BL & BL \\
        2002dl & CN & CN & CL \\
        2003gt & CN & CN & CL \\
        2011iv & CN & CN & CL \\
        ASASSN-14hu & SS & CN & CN \\
        2003gn & CL & BL & BL \\
        PS1-14ra & CL & CL & CN \\
        PTF14w & CL & CL & CN \\
        2007on & CL & CL & CN \\
        2008ec & CL & CL & CN \\
        2008fl & CL & CL & CN \\
        2011jh & CL & CL & CN \\
    \enddata
    \end{deluxetable}

\subsection{\polinplot Groups}
\label{sec:polin-plot-discussion}

\begin{figure}[ht]
    \centering
    \includegraphics[scale=1.3]{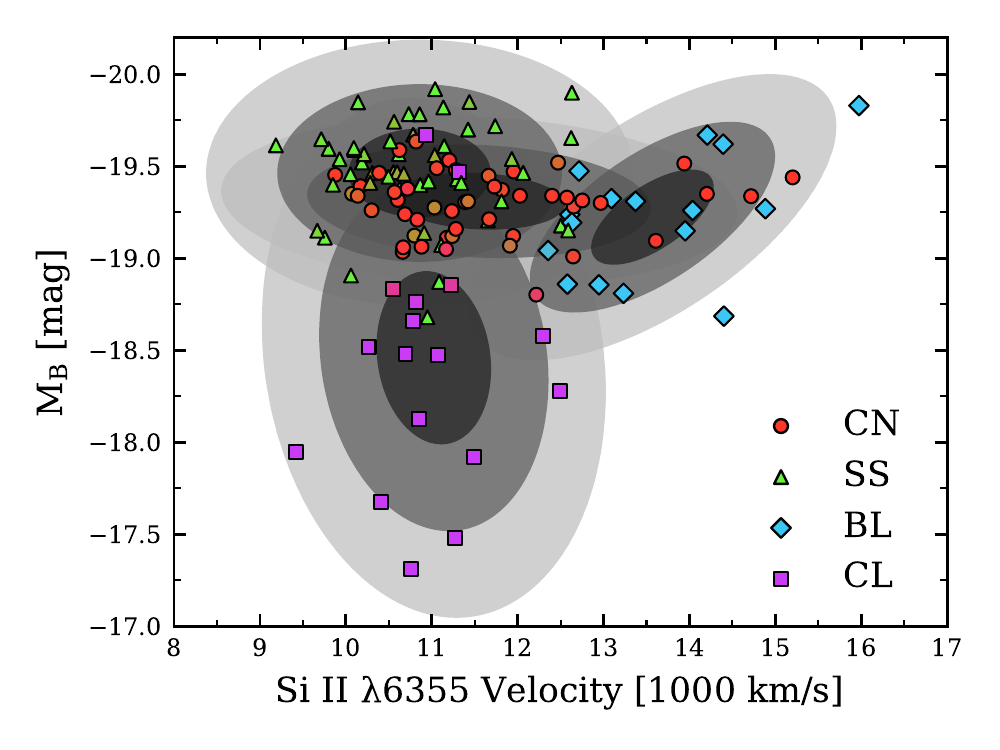}
    \caption{The \polinplot diagram of the 4-D GMM analysis from
             \autoref{sec:4D-GMM}. The contours indicate covariance (up to
             $3\sigma$) of group membership projected into [$M_B$, \vsi] space.
             We see that the Main group consists of both CN and SS objects and
             that the Fast group is separated into both BL and CN objects. This
             separation implies there is a dispersion in the relationship
             between \vsi and pEW(\Rline). As expected, the Dim group almost
             exclusively contains CL objects.}
    \label{fig:m_v_GMM_m_v_p5_p6}
\end{figure}

Our analysis does not show the clear dichotomy in [$M_B$, \vsi] space found in
\citet{polin19hedet} when we include the CSP I+II samples
in addition to the \citet{Zheng_IaLCIII18} sample. In particular, their result seems to be
due to not correcting for reddening in the host, when plotting \BV. The only
intrinsically red objects are the Dims/CLs. When all four parameters are
accounted for in the GMM, as shown in \autoref{fig:m_v_GMM_m_v_p5_p6}, we see that
the Main group consists primarily of CN and SS objects, the Fast group is
dominated by BL objects with only a few CN objects, and the Dim group is again
made up primarily of CL objects. The Fast objects can indeed be divided using
their pEW properties. Therefore, the Fast group SNe do not need to stem from
sub-Chandrasekhar explosions, and additional information is required to more
concretely determine the underlying physics involved.

If we assume the Main group to be the union of both SS and CN objects, we do
find that, in comparing \autoref{fig:m_v_GMM_polin} with the modified \polinplot
diagram  presented in \autoref{fig:m_v_GMM_m_v_p5_p6}, the projected covariances into [$M_B$,
\vsi] space show significant qualitative change. The BL contours that are
associated with the Fast group are noticeably smaller, meaning it is a more
well-constrained grouping with less uncertainty between the BLs and other
groups. The Main group contains both CN and SS objects so it is important to
determine if there are other distinctions between CN and SS objects
that can be related to variations in the progenitors or explosion mechanisms.

From \autoref{fig:m_v_BV_GMM_polin_COMP} we see that the 4-D grouping system
projected onto [$M_B$, \vsi] space does not sort SNe by \BV color. We find
that \BV is mostly distinct only for Dims, and the Fast group color is
bluer than Dims, regardless of whether the object is a CN or BL SN. More
work must be done to further sub-classify these SNe~Ia.

\subsection{Branch Group Relation with $s_{BV}$}

\begin{figure}[ht]
    \centering
    \includegraphics[scale=1.1]{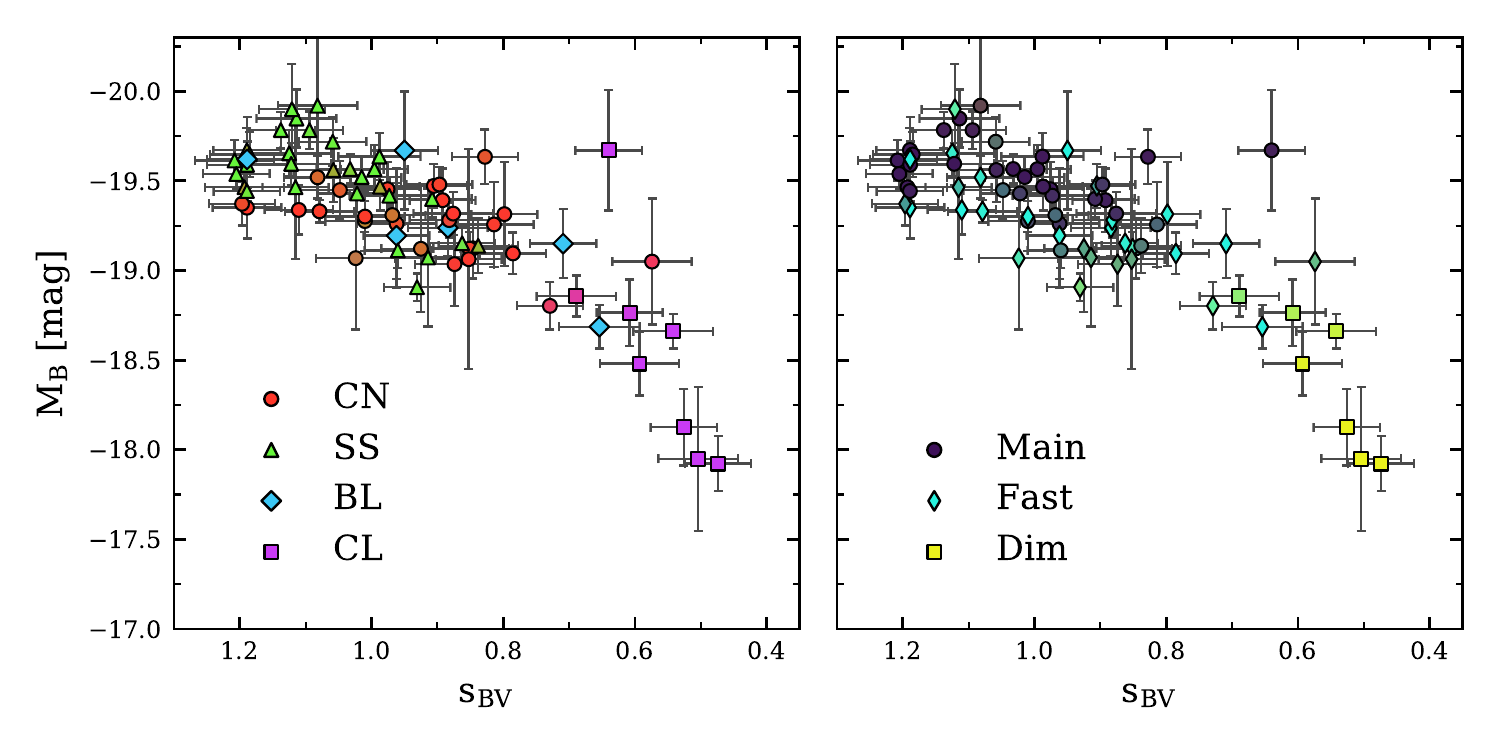}
    \caption{Illustration of the Phillips relation as $M_B$ versus $s_{BV}$.
             Points are color-coded by the 4-D Branch groups (left panel) and
             the \polinplot groups (right panel).}
    \label{fig:m_sbv_GMM_m_v_p5_p6_GMM_polin}
\end{figure}

\begin{figure}[ht]
    \centering
    \includegraphics[scale=1.3]{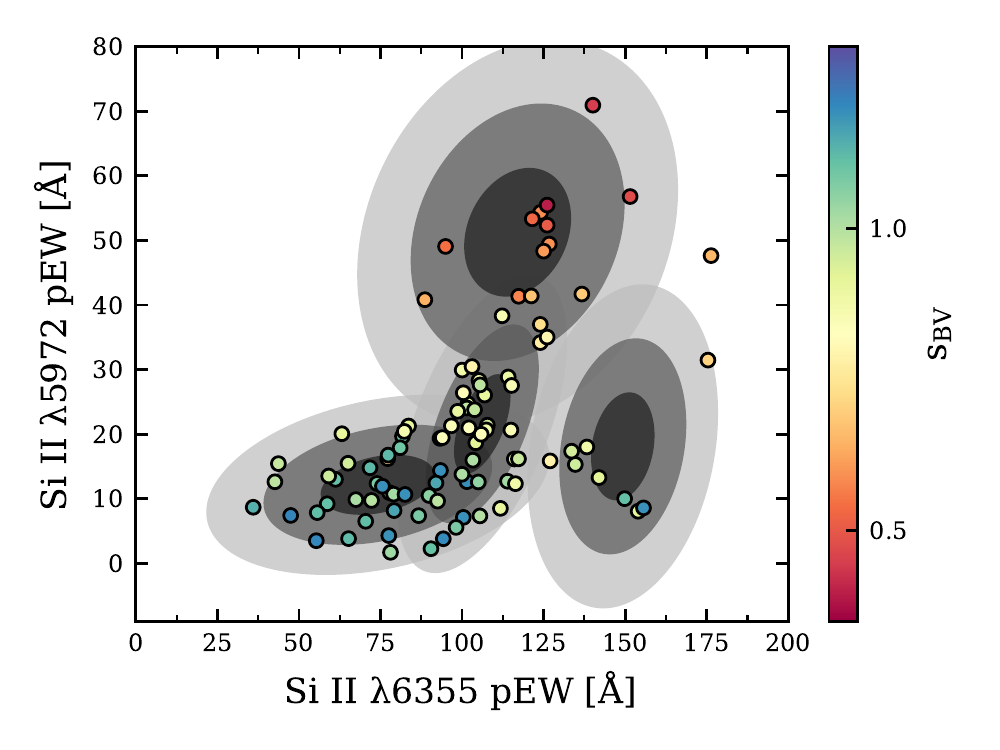}
    \caption{Branch diagram of the CSP I+II samples coded for color stretch
             parameter $s_{BV}$. The contours are exactly those from
             \autoref{fig:m_v_p5_p6_GMM_m_v_p5_p6}, indicating our defined 4-D
             Branch groups.}
    \label{fig:p5_p6_sBV_GMM_m_v_p5_p6}
\end{figure}

\begin{figure}[ht]
    \centering
    \includegraphics[scale=1.1]{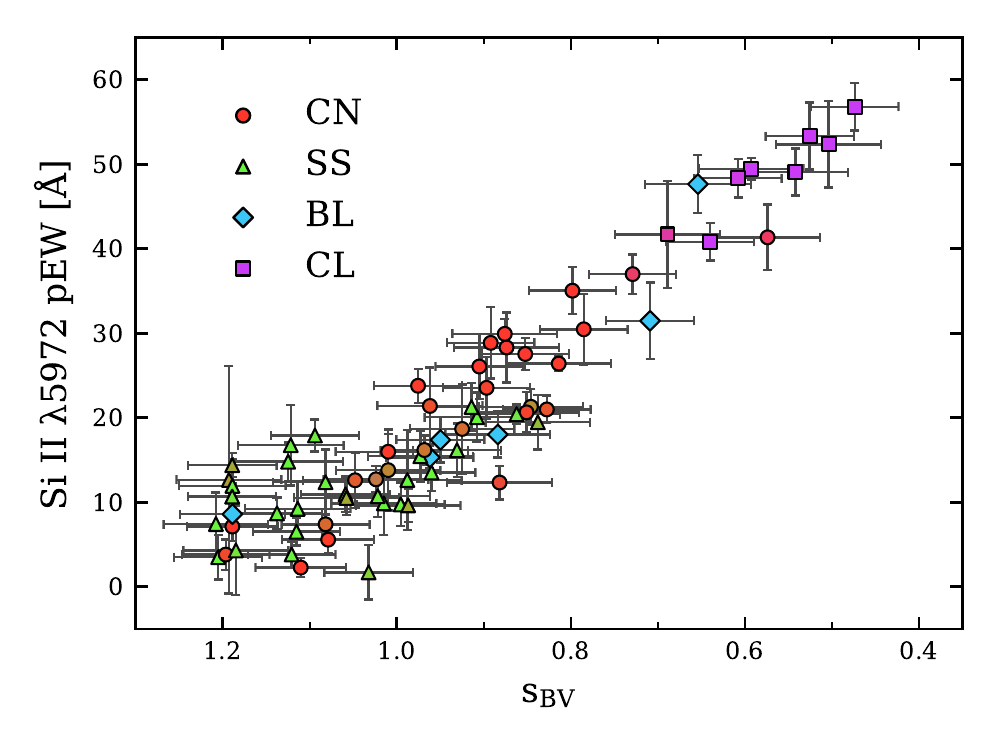}
    \caption{pEW(\Bline) versus $s_{BV}$ of the CSP I+II samples colored by
             the 4-D Branch groups. We see a linear trend for SNe with $s_{BV}
             \lesssim 1.0$.}
    \label{fig:p5_sbv_GMM_m_v_p5_p6}
\end{figure}

\autoref{fig:m_sbv_GMM_m_v_p5_p6_GMM_polin} shows the Phillips/Burns relation
color-coded by Branch group (left panel) and by \polinplot group (right panel).
Both the CL and Dims are intrinsically dim, and there is much overlap between
them, as seen in \autoref{sec:polin-plot-discussion}. The Branch diagram coded
for $s_{BV}$ is displayed in \autoref{fig:p5_p6_sBV_GMM_m_v_p5_p6} and it shows
that $s_{BV}$ does not distinguish between CN and BL SNe with high values of
\vsi as is seen in \autoref{fig:m_v_GMM_m_v_p5_p6}. It will be interesting to
understand what causes this variation to better improve the use of SNe~Ia as
cosmological probes.

In \autoref{fig:m_sbv_GMM_m_v_p5_p6_GMM_polin} there appears to be a transition
region of SS into CN that corresponds to a transition of the Main group into
the Fast group. It could be that there is some underlying parameterization
involved that better explains the dispersion in this Phillips/Burns relation,
such as a dependency upon \polinplot group membership or \vsi. This interesting
behavior is left to be studied in future work.

\autoref{fig:p5_sbv_GMM_m_v_p5_p6} shows that a Phillips-like relation exists for
pEW(\Bline) and the color stretch parameter, $s_{BV}$ \citep{Burns_CSP18}. In
other words, pEW(\Bline) acts as a stand-in for $M_B$ (see
\Autoref{{fig:m_p5_p6_GMM_m_p5_p6}, {fig:m_p5_GMM_m_p5_p6}}).
\autoref{fig:p5_sbv_GMM_m_v_p5_p6} also shows the 4-D Branch group placement for
this relation. At high $s_{BV}$ ($s_{BV} \gtrsim 1.0$), little correlation is
found between pEW(\Bline) and $s_{BV}$, which is expected because the
dependency of $s_{BV}$ on $M_B$ decreases for brighter SNe (seen in
\autoref{fig:m_sbv_GMM_m_v_p5_p6_GMM_polin}). However, a linear trend is found for
$s_{BV} \lesssim 1.0$.

\clearpage

\section{Conclusions}

We have shown that a 4-D Gaussian mixture model (GMM) analysis of [$M_B$, \vsi,
pEW(\Bline), pEW(\Rline)] with $n=4$ components yields robust groupings that
strongly identify the four Branch groups: core-normals (CN), shallow-silicons
(SS), broad-lines (BL), and cools (CL). We have shown that there seems to be
a strong correlation between $M_B$ and pEW(\Bline), and because of this we
suggest that the quantities pEW(\Rline), pEW(\Bline), and \vsi can be used to
approximate this 4-D model with only spectroscopic quantities. Furthermore, the
3-D GMM makes the BL group nearly distinct; it seems reasonable that this
subclass of SNe would be distinct from the rest of the SNe Ia as
suggested previously by \citet{WWFZZ13}. One possibility is that the high
velocity of the photosphere is produced by a shell, such as that produced by a
pulsating delayed detonation \citep{HKWPSH96,dessart11fe14}.

The original \polinplot diagram was interpreted to separate SNe~Ia into two
groups based on $M_B$ and \vsi that likely corresponded to Chandrasekhar-mass
explosions and sub-Chandrasekhar explosions \citep{polin19hedet}. We find that
this is an incomplete description; rather than a clear dichotomy in [$M_B$,
\vsi] space, there are three connected groups. Ultimately we find that SN
subtypes are better-delineated by the Branch groups using GMMs. We define these
three \polinplot groups as: the Main group consisting of CNs and SSs, the Fast group
consisting of BLs and a subset of CNs, and the Dim group that consists of CLs.
It is shown that Fast objects can be divided using their Branch group
membership (\ion{Si}{2} pEWs), and this division may aid in predicting their
underlying explosion mechanisms. We find that this separation is generally
unexplained by the color stretch parameter $s_{BV}$, and so future work must be
done to explain this dispersion.

Open access to each of these GMMs is provided at
\url{https://github.com/anthonyburrow/SNIaDCA}. In future work we plan to use a
tool such as principal component analysis to determine just how much pEW(\Bline)
is captured in $M_B$ and how much pEW(\Rline) is captured in \vsi and other
covariances including \BV and $s_{BV}$ as well as extinction properties such as
color excess $E(B-V)$ and $R_V$.

\section{Acknowledgments}
The work of the CSP-II has been generously supported by the NSF under
grants AST-1008543, AST-1613426, AST-1613455, AST-1613472,  and in
part by a Sapere Aude Level 2 grant funded by the Danish Agency for
Science and Technology and Innovation  (PI M.S.).
AB and EB were supported in part by NASA grant 80NSSC20K0538.
 L.G. was funded by the European Union's Horizon 2020 research and
 innovation programme under the Marie Sk\l{}odowska-Curie grant
 agreement No. 839090. This work has been partially supported by the
 Spanish grant PGC2018-095317-B-C21 within the European Funds for
 Regional Development (FEDER). M.S. is supported by generous grants
 (13261 and 28021) from VILLUM FONDEN, and also by  a project grant
 (8021-00170B) from the Independent Research Fund Denmark. PJB, KK, and NBS
 gratefully acknowledge the support of the George P. and Cynthia Woods
 Mitchell Institute for Fundamental Physics and Astronomy. We also
 thank the Mitchell Foundation for their sponsorship of the Cook's
 Branch Workshop on Supernovae where much of this science was
 discussed.

\software{SNooPy \citep{Burns_CSP11}, Spextractor,
GPy \citep[version 1.9.9,][]{gpy2014},
scikit-learn \citep[version 0.22.2,][]{scikit},
NumPy \citep[version 1.18.2,][]{numpy1,numpy2},
Matplotlib \citep[version 3.2.1,][]{matplot}}

\clearpage

\appendix

\section{Inclusion of Errors to GP Kernel}
\label{sec:appx-GP}

In this section we briefly describe most major changes to \texttt{Spextractor}
we invoke pertaining to spectrum fitting using Gaussian process regression
(GPR) that are used in calculating line velocities and pseudo-equivalent
widths. The background information given paraphrases a more detailed
explanation provided in \citet{Murphy2012}.

A Gaussian process is used to infer a distribution over functions $p(f |
\mathbf{X}, \mathbf{y})$ given some observed input set $\mathbf{X}$ and output
set $\mathbf{y}$ such that $y_i = f(x_i)$ for $y_i \in \mathbf{y}$ and
$x_i \in \mathbf{X}$. In this paper, for fitting spectra with a GPR,
$\mathbf{y}$ is the set of flux measurements, and $\mathbf{X}$ is the
corresponding wavelengths at which these measurements were taken. The Gaussian
process is defined to assume that $p(f(x_1), ..., f(x_N))$ is jointly Gaussian
for an arbitrary set of inputs $x_1, ..., x_N$ for $N$ observation points in $
\mathbf{X}$ and therefore in $\mathbf{y}$. This joint Gaussian distribution has
mean $\mathbf{\mu}(\mathbf{x})$ and covariance $\mathbf{K}(\mathbf{x})$ given
by $K_{ij} = \kappa(x_i, x_j)$, where $\kappa$ is the positive definite kernel
function for which we choose the Mat\'ern 3/2  covariance function
\citep{RasWil06} as does the original \texttt{Spextractor} code. The functional
distribution is then normally distributed as
\begin{equation}
\begin{pmatrix}
\mathbf{f} \\
\mathbf{f}*
\end{pmatrix}
=
\mathcal{N}\left(
\begin{pmatrix}
\mathbf{\mu} \\
\mathbf{\mu}_*
\end{pmatrix}
,
\begin{pmatrix}
\mathbf{K} & \mathbf{K}_* \\
\mathbf{K}_*^\mathsf{T} & \mathbf{K}_{**}
\end{pmatrix}
\right),
\end{equation}
where $\mathbf{f}_*$ are functional outputs of $\mathbf{X}$, and $\mathbf{K} =
\kappa(\mathbf{X}, \mathbf{X})$, $\mathbf{K}_* = \kappa(\mathbf{X},
\mathbf{X}_*)$, and $\mathbf{K}_{**} = \kappa(\mathbf{X}_*, \mathbf{X}_*)$,
where $\mathbf{X}_*$ is some test (or prediction) input set. The output values
$\mathbf{f}_*$ are then predicted with a GPR as mean values and associated
variances for each test point. In this work we modify \texttt{Spextractor} such
that $\mathbf{X}_* \neq \mathbf{X}$ to allow a representation of the variance
between observed data points. More specifically, $\mathbf{X}_*$ was selected as
a uniform distribution of 2,000 values that spanned the given spectrum.

If there are independently and identically distributed uncertainties
$\epsilon_i$ \citep[see, for example,][]{Murphy2012} in the observed output
such that
\begin{equation}
y_i = f(x_i) + \epsilon_i,
\end{equation}
where in general each $\epsilon_i$ may not be equal to one another, we may
write the distribution as
\begin{equation}
\begin{pmatrix}
\mathbf{\mathbf{y}} \\
\mathbf{f}*
\end{pmatrix}
=
\mathcal{N}\left(
\begin{pmatrix}
\mathbf{\mu} \\
\mathbf{\mu}_*
\end{pmatrix}
,
\begin{pmatrix}
\mathbf{K}_y & \mathbf{K}_* \\
\mathbf{K}_*^\mathsf{T} & \mathbf{K}_{**}
\end{pmatrix}
\right),
\end{equation}
where $\mathbf{K}_y$ is the covariance matrix defined by $\mathbf{K}_y \equiv
\mathbf{K} + (\sigma^2)^\mathsf{T} I_N$ with $\sigma^2$ being a vector of
associated output uncertainties and $I_N$ the $N \times N$ identity. We
therefore add flux uncertainties $\sigma_{flux}^2$ in quadrature to the kernel
when flux uncertainties were provided.

\bibliography{msrefs}

\end{document}